\documentclass[aps,
	twocolumn,
	superscriptaddress,
	10pt]{revtex4-1}

\usepackage{amsmath}
\usepackage{graphicx}
\usepackage{bm}
\usepackage{gensymb,textcomp}
\usepackage{times}
\usepackage{hyperref}
\usepackage{subfigure}
\usepackage{color}
\hypersetup{pdfpagemode=UseNone}

\newcommand{\nhat}{\hat{n}}
\newcommand{\xhat}{\hat{x}}
\newcommand{\yhat}{\hat{y}}
\newcommand{\zhat}{\hat{z}}
\newcommand{\khat}{\hat{k}}

\newcommand{\bk}{\mathbf{k}}

\newcommand{\bE}{\mathbf{E}}

\renewcommand{\t}[1]{\text{#1}}
\newcommand{\ee}{\text{ee}}

\renewcommand{\Re}{\operatorname{Re}}
\renewcommand{\Im}{\operatorname{Im}}

\begin{document}

\title{Transmission magnitude and phase control for
polarization-preserving reflectionless metasurfaces}

\author{Do-Hoon Kwon}
\email{dhkwon@umass.edu}
\affiliation{Department of Electrical and Computer Engineering,
University of Massachusetts Amherst, Amherst, Massachusetts 01003, USA}
\author{Grigorii Ptitcyn}
\affiliation{Department of Electronics and Nanoengineering,
Aalto University, P.O. Box 15500, 00076 Aalto, Finland}
\author{Ana D\'iaz-Rubio}
\affiliation{Department of Electronics and Nanoengineering,
Aalto University, P.O. Box 15500, 00076 Aalto, Finland}
\author{Sergei A. Tretyakov}
\affiliation{Department of Electronics and Nanoengineering,
Aalto University, P.O. Box 15500, 00076 Aalto, Finland}

\begin{abstract}
For transmissive applications of electromagnetic metasurfaces,
an array of subwavelength Huygens' meta-atoms are typically
used to eliminate reflection and
achieve a high transmission power efficiency together
with a wide transmission phase coverage. We show that the
underlying principle of low reflection and 
full control over transmission
is asymmetric scattering into the specular reflection and
transmission directions that results from a superposition
of symmetric and anti-symmetric scattering components, with
Huygens' meta-atoms being one example configuration.
Available for oblique illumination in TM polarization,
a meta-atom configuration comprising
normal and tangential electric polarizations is presented,
which is capable of reflectionless, full-power transmission
and a $2\pi$ transmission phase coverage as well as full absorption.
For lossy metasurfaces, we show that a complete phase coverage
is still available for reflectionless designs for any value of
absorptance. Numerical examples in the microwave
and optical regimes are provided.
\end{abstract}

\maketitle

\section{Introduction}
\label{sec:introduction}

Presented as two-dimensional equivalents of volumetric
metamaterials~\cite{holloway_ieeemap2012}, metasurfaces have
attracted a significant interest in recent years. Metasurfaces
are typically realized as a doubly periodic array of small
polarizable particles over a subwavelength thickness. With
a distinct advantage of low loss over volumetric metamaterials,
a wide range of novel reflection, transmission, and absorption
applications ranging from microwave to optical frequencies
have been
reported~\cite{kildishev_science2013,yu_natmater2014,meinzer_natphoton2014,
glybovski_physrep2016,chen_repprogphys2016}.

There are a host of metasurface applications in the transmission
mode. Based on the phased array antenna principle~\cite{hansen2009} and
the generalized law of refraction~\cite{yu_science2011},
a linear gradient of transmission phase imparted on the
transmitted wave can bend an incident beam or plane wave in an
anomalous 
direction~\cite{pfeiffer_prl2013,aieta_nanolett2012b,pfeiffer_nanolett2014,
yu_lasphotonrev2015}. A flat focusing lens is obtained by
spatially nonlinear transmission phase
distributions~\cite{monticone_prl2013,lin_science2014,
wang_advopticalmat2015,khorasaninejad_science2016}.
Local modulation of transmission amplitude and/or phase leads
to holograms~\cite{ni_natcommun2013,huang_natcommun2013,
arbabi_natnanotech2015}. As polarization transformers,
thin metasurfaces can replace electrically
thick wave plates by assigning distinct transmission phases to
two orthogonal polarization 
components~\cite{zhao_natcommun2012,pfeiffer_ieeejmtt2013,
grady_science2013}.

A high transmission magnitude toward unity and a
wide transmission phase range toward $2\pi$ are highly
desirable for high-efficiency operation of transmitted
wave shaping. At optical frequencies, a single layer of
plasmonic or dielectric resonant particles are commonly adopted
for transmissive metasurfaces, owing to relative ease of
fabrication. It was recognized that an infinitesimally
thin layer of electrically polarizable particles support
a complete $2\pi$ range for the cross-polarized transmission
phase, but not for the co-polarized transmission.
Accordingly, formalization of the generalized laws of
reflection and refraction as well as their experimental
demonstrations were performed for the cross-polarized
components scattered by $V$-shaped optical antennas under linearly
polarized illuminations~\cite{yu_science2011,aieta_nanolett2012},
albeit at a low cross-polarized transmission efficiency.
It was later revealed that the maximum cross-coupled power
efficiency is 25\%~\cite{monticone_prl2013}.
Another approach to achieve a $2\pi$ transmission phase
coverage is to exploit the Pancharatnam-Berry phase together with
circularly polarized illuminations. For the transmitted
circularly polarized wave of the opposite handedness,
the phase can be adjusted simply by rotating the principal 
axes of a wave-plate
element~\cite{lin_science2014,zheng_natnanotech2015,
huang_natcommun2013}. Still, this approach involves cross-polarized
transmitted waves and a completely reflectionless operation
is not possible.

For polarization-preserving applications to date,
reflectionless metasurfaces with a complete $2\pi$
transmission phase coverage are based on Huygens'
meta-atoms~\cite{pfeiffer_prl2013,yu_lasphotonrev2015,
shalaev_nanolett2015,decker_advoptmat2015,chong_acsphoton2016}.
A Huygens' meta-atom is designed such that an orthogonal set
of tangential electric and magnetic dipole moments are induced
upon external plane-wave excitation. When the two induced
dipoles satisfy a balanced condition, full transmission
and zero reflection can be obtained for lossless cases
with the transmission phase that can be designed to have any
value within a complete $2\pi$ range. In the microwave regime,
short conductor traces are typically used for electric dipoles.
For equivalent magnetic dipoles, either metallic 
ring resonators~\cite{pfeiffer_prl2013}
or conductor traces in a multi-layer dielectric
substrate~\cite{pfeiffer_ieeejmtt2013,wong_ieeejmtt2015,epstein_josab2016}
are utilized.
{\color{red}Utilizing a combination of continuous and discrete
printed multi-layer conductor traces for realizing Huygens' meta-atoms,
microwave transmissive metamaterials for lenses, beam deflectors,
and vortex beam generators with efficiencies as high as 91\% have
been demonstrated~\cite{cai_advoptmat2017,luo_prappl2017,cai_prappl2017}.} 
In the optics regime, a dielectric resonator meta-atom can
be designed to support both electric and equivalent magnetic
polarization currents at the same 
wavelength~\cite{decker_advoptmat2015,yu_lasphotonrev2015,
shalaev_nanolett2015}. It is challenging to design induced
electric and magnetic polarizations to satisfy
a balanced amplitude and phase relation at the same frequency
due to strong mutual coupling between them.
Furthermore, the bandwidth of the magnetic resonance tends to be
narrower than that of the electric one.
At microwave frequencies, the length and shape of a thin conductor
wire may be designed to strike a balance between the induced
electric and equivalent magnetic dipole moments. This approach 
has been demonstrated with a circularly polarized Huygens spiral
particle in \cite{niemi_ieeejap2013}.

In this paper, a fundamental principle behind full power
transmission with a complete phase coverage is investigated
for polarization-preserving metasurfaces.
It is shown that generation of an
anti-symmetric scattering component by the
induced polarizations and its destructive interference with the
symmetric scattering component in nullifying the total reflected wave
is the key design principle. In Huygens' metasurfaces, a tangential
magnetic dipole moment provides the necessary anti-symmetric scattering.
However,
it is not the only possible source of anti-symmetric scattering.
It has been recently shown that the effect of a tangential
magnetic polarization can be equivalently generated by a spatially
varying normal electric polarization~\cite{albooyeh_prb2017}.
Taking advantage of this equivalence,
a meta-atom configuration comprising entirely of electric
dipole moments that is capable of full transmission with a $2\pi$
phase coverage is presented. The new reflectionless metasurface
configuration is available for oblique illuminations in the
TM polarization. Here, the anti-symmetric scattering is provided
by an electric dipole that is polarized normal to the metasurface.

{\color{red}When a combination of tangential and normal
induced electric dipoles are realized with a tilted electric dipole
meta-atom, the identical frequency dispersion of the
two orthogonal dipole moments permits 
\emph{dispersionless} full transmission
at a \emph{fixed} oblique incidence angle. Furthermore,
the resonance of a dipole meta-atom can be tuned in order to
achieve a full transmission phase coverage.
It is known that metallic gratings having an array
of narrow slits or apertures allow dispersionless broadband
extraordinary transmission
under a TM-polarized illumination at some
fixed oblique angle~\cite{huang_prl2010,alu_prl2011,fan_advmat2012}.
The underlying physics of this broadband full transmission
is an impedance match between an oblique TM-polarized illumination
and a propagating mode inside a slit~\cite{alu_prl2011}.
In contrast, the dispersionless transmission property
of a tilted dipole array in this study is of geometrical nature
of the dipole pattern null directions. This allows the frequency
dispersion of a dipole meta-atom to be exploited for transmission
phase synthesis.}

Furthermore, the inevitable effect of absorption in
practical metasurface realizations on the transmission phase
coverage is discussed.
It is revealed that a complete $2\pi$ range of transmission
coverage is still available regardless of the amount of loss
for reflectionless designs.
Finally, illustrative uniform metasurface design 
examples are provided and analyzed.
As a microwave example, a tilted thin conducting strip dipole
array is shown for providing full transmission and a complete phase
coverage. As its lossy variant, an impedance-loaded
strip dipole array is
analyzed for full phase coverage as well as for perfect absorption. 
For an example in the optical
regime, high transmission and full phase coverage by an
all-dielectric metasurface composed of dielectric bars
of rectangular cross section are demonstrated.

In the following time-harmonic analysis at an angular frequency
$\omega$, an $e^{j\omega t}$ time dependence is assumed 
and suppressed.

\section{TM-mode reflection and transmission}
\label{sec:model}

\begin{figure}[t]
 \centering
 \includegraphics{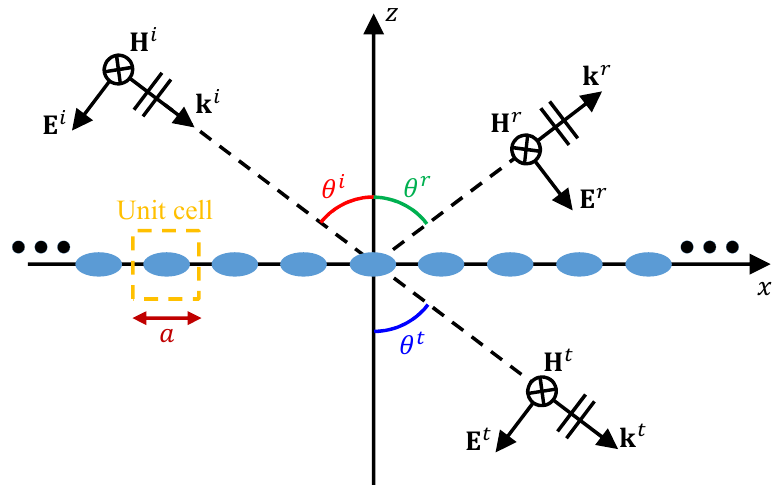}
 \caption{Reflection and transmission of a TM-polarized plane
wave by a metasurface in the $xy$-plane situated in free space.}
 \label{fig:tm_refl_trans}
\end{figure}

We treat a uniform (non-gradient), single-layer metasurface
in free space. As will be shown, lossless metasurfaces supporting
electric polarizations only are capable of full-power transmission
in the TM polarization. Figure~\ref{fig:tm_refl_trans} illustrates
a planar metasurface in the $xy$-plane illuminated by a TM-polarized
plane wave propagating in the $xz$-plane with an angle of incidence
$\theta^\t{i}$. The unit-cell
dimensions are $a$ and $b$ in the $x$- and $y$-axis directions,
respectively.
The dimensions are set such that only the fundamental Floquet
mode fields propagates away from the metasurface. Away from the
metasurface where all evanescent higher-order Floquet mode
waves vanish, the incident, reflected, and transmitted electric
fields can be written as
\begin{equation}
\bE^\t{p}=\bE^\t{p}_0e^{-j(k^\t{p}_xx+k^\t{p}_zz)};\ \ 
\t{p}=\t{i},\t{r},\t{t},
\label{def:E-fields}
\end{equation}
where the superscripts `i,' `r,' and `t' denote the incident,
reflected, and transmitted fields, respectively, and
$\bE^\t{p}_0$ denotes the corresponding E-field vector amplitude.
The associated wave vectors in the $xz$-plane are given by
$(k^\t{i}_x,k^\t{i}_z)=(k\sin\theta^\t{i},-k\cos\theta^\t{i})$,
$(k^\t{r}_x,k^\t{r}_z)=(k\sin\theta^\t{r},k\cos\theta^\t{r})$,
and $(k^\t{t}_x,k^\t{t}_z)=(k\sin\theta^\t{t},-k\cos\theta^\t{t})$,
with $k$ denoting the free-space wavenumber. All magnetic
fields are $y$-polarized. For a uniform metasurface, all three
angles are the same $(\theta^\t{i}=\theta^\t{r}=\theta^\t{t})$.

For metasurfaces that do not scatter cross-polarized fields,
the transmission and reflection in Fig.~\ref{fig:tm_refl_trans}
can be described using a two-port network in terms of $S$-parameters,
with plane-wave ports
1 and 2 defined in positive- and negative-$z$ half spaces.
With the phase reference planes for both ports set to $z=0$,
the reflection coefficient $r=S_{11}$ and the transmission
coefficient $t=S_{21}$ are defined as the ratios of 
tangential E-field amplitudes as
\begin{equation}
r=\frac{\xhat\cdot\bE^\t{r}_0}{\xhat\cdot\bE^\t{i}_0}
=\frac{E^\t{r}_{0x}}{E^\t{i}_{0x}},\ \
t=\frac{\xhat\cdot\bE^\t{t}_0}{\xhat\cdot\bE^\t{i}_0}
=\frac{E^\t{t}_{0x}}{E^\t{i}_{0x}}.
\label{def:r_and_t}
\end{equation} 

\begin{figure}[t]
 \centering
 \includegraphics*{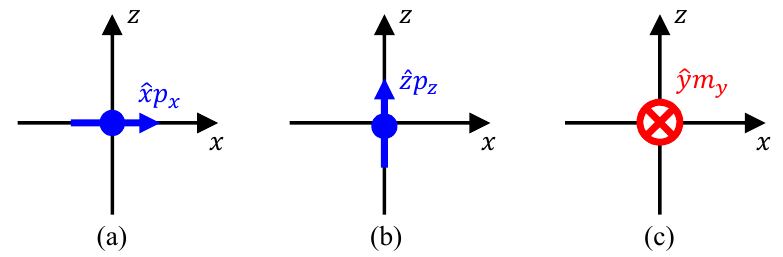}
 \caption{Three elemental dipole meta-atoms for TM-polarization
reflection and scattering. (a) An $x$-directed electric dipole.
(b) A $z$-directed electric dipole. (c) A $y$-directed magnetic
dipole.}
 \label{fig:elemental_dipoles}
\end{figure}

There are three elemental dipole meta-atoms that scatter
TM-polarized fields, which are assumed to be excited by
the incident wave.
They are two orthogonal electric dipoles in the
plane of incidence (the $xz$-plane)
and one magnetic dipole that is directed
normal to the plane of incidence, as shown in
Fig.~\ref{fig:elemental_dipoles}. Each point dipole is positioned
at the coordinate origin. A general meta-atom can
be represented as a superposition of these three dipoles.
The scattering characteristics of the elemental dipoles
are discussed next.

\subsection{Symmetric scattering}
\label{sec:symmetric}

A planar array of the tangential electric dipole $\xhat p_x$ in
Fig.~\ref{fig:elemental_dipoles}(a) radiates plane waves
symmetrically into $z\to\pm\infty$. The scattered E-field
amplitudes, $\bE^{\t{s},\t{r}}_0$ and $\bE^{\t{s},\t{t}}_0$,
in the reflection and transmission directions, respectively,
are given by
\begin{eqnarray}
\bE^{\t{s},\t{r}}_0 &=& \frac{j\omega\eta}{2S\cos\theta^\t{i}}
\khat^\t{r}\times\khat^\t{r}\times\xhat p_x,
\label{def:Esr0_px}\\
\bE^{\t{s},\t{t}}_0 &=& \frac{j\omega\eta}{2S\cos\theta^\t{i}}
\khat^\t{t}\times\khat^\t{t}\times\xhat p_x,\label{def:Est0_px}
\end{eqnarray}
where $\eta\approx 377~\Omega$ is the free-space intrinsic
impedance and $S=ab$ represents the unit-cell area in the
$xy$-plane. Also, $\khat^\t{r}=\bk^\t{r}/k$ and 
$\khat^\t{t}=\bk^\t{t}/k$ are the unit vectors in their respective
propagation directions.
The tangential components of $\bE^{\t{s},\t{r}}_0$ and
$\bE^{\t{s},\t{t}}_0$ are the same (symmetric). Denoting
this value by $E^{\t{s},\t{s}}_{0x}$, we find
\begin{equation}
E^{\t{s},\t{s}}_{0x}=\xhat\cdot\bE^{\t{s},\t{r}}_0
=\xhat\cdot\bE^{\t{s},\t{t}}_0=-\frac{j\omega\eta p_x}{2S}
\cos\theta^\t{i}.
\label{def:Ess0x}
\end{equation}
These scattered field amplitudes could be cast in terms of
the surface-averaged electric polarization current density
$J_x=j\omega p_x/S$.

\subsection{Anti-symmetric scattering}
\label{sec:anti-symmetric}

Both a normal electric dipole [Fig.~\ref{fig:elemental_dipoles}(b)]
and a tangential magnetic dipole
[Fig.~\ref{fig:elemental_dipoles}(c)] radiate anti-symmetrically
into $z\to\pm\infty$. First, the $z$-directed electric
dipole meta-atom generates scattered plane waves in the
$\khat^\t{r}$, $\khat^\t{t}$ directions with 
the E-field amplitudes given by
\begin{eqnarray}
\bE^{\t{s},\t{r}}_0 &=& \frac{j\omega\eta}{2S\cos\theta^\t{i}}
\khat^\t{r}\times\khat^\t{r}\times\zhat p_z,
\label{def:Esr0_pz}\\
\bE^{\t{s},\t{t}}_0 &=& \frac{j\omega\eta}{2S\cos\theta^\t{i}}
\khat^\t{t}\times\khat^\t{t}\times\zhat p_z.
\label{def:Est0_pz}
\end{eqnarray}
The $x$-components of these two vectors
are different by sign (anti-symmetric).
Denoting the anti-symmetric component by $E^{\t{s},\t{a}}_{0x}$,
we find
\begin{equation}
E^{\t{s},\t{a}}_{0x}=\xhat\cdot\bE^{\t{s},\t{r}}_0=
-\xhat\cdot\bE^{\t{s},\t{t}}_0=\frac{j\omega\eta p_z}{2S}\sin\theta^\t{i}.
\label{def:Esa0x_pz}
\end{equation}

Next, a planar array of the $y$-directed magnetic dipole meta-atom
in Fig.~\ref{fig:elemental_dipoles}(c) creates 
the E-field amplitudes for the scattered plane waves given by
\begin{eqnarray}
\bE^{\t{s},\t{r}}_0 &=& \frac{j\omega}{2S\cos\theta^\t{i}}
\khat^\t{r}\times\yhat m_y,
\label{def:Esr0_my}\\
\bE^{\t{s},\t{t}}_0 &=& \frac{j\omega\eta}{2S\cos\theta^\t{i}}
\khat^\t{t}\times\yhat m_y.
\label{def:Est0_my}
\end{eqnarray}
The anti-symmetric $x$-components can be written as
\begin{equation}
E^{\t{s},\t{a}}_{0x}=\xhat\cdot\bE^{\t{s},\t{r}}_0=
-\xhat\cdot\bE^{\t{s},\t{t}}_0=-\frac{j\omega m_y}{2S}.
\label{def:Esa0x_my}
\end{equation}
The field amplitudes can also be written in terms of the
surface-averaged polarization current densities,
$J_z=j\omega p_z/S$ and $M_y=j\omega m_y/S$.

\subsection{Transmission and reflection coefficients for total fields}
\label{sec:total_scattering}

Infinitely thin metasurfaces can support tangential
electric polarizations only and the resulting symmetric scattering
is the reason for low transmission power efficiencies for
polarization-preserving designs~\cite{monticone_prl2013}.
For Huygens' metasurfaces, the asymmetric scattering enabled
by a combination of orthogonally directed tangential electric
and magnetic dipoles allows full
transmission~\cite{yu_lasphotonrev2015,decker_advoptmat2015}.
In order to support an equivalent magnetic dipole realized
using a circulating electric polarization, a Huygens' metasurface
cannot have a vanishingly small thickness.

We note that the anti-symmetric scattering may be provided
by a normal electric dipole rather than a tangential magnetic
dipole. Regardless of the origin of anti-symmetric radiation,
the total tangential E-field components for the reflected
and transmitted waves can be written as a superposition as
\begin{equation}
E^\t{r}_{0x}=E^{\t{s},\t{s}}_{0x}+E^{\t{s},\t{a}}_{0x},\ \ 
E^\t{t}_{0x}=E^\t{i}_{0x}+E^{\t{s},\t{s}}_{0x}-E^{\t{s},\t{a}}_{0x},
\end{equation}
where the symmetric component $E^{\t{s},\t{s}}_{0x}$
is given by (\ref{def:Ess0x}). The anti-symmetric component
$E^{\t{s},\t{a}}_{0x}$ is given by (\ref{def:Esa0x_pz}) if
$\zhat p_z$ is used and by (\ref{def:Esa0x_my}) if
$\yhat m_y$ is used. Then, we can write the reflection and
transmission coefficients as
\begin{eqnarray}
r &=& \frac{E^\t{r}_{0x}}{E^\t{i}_{0x}}=r_\t{s}+r_\t{a};\ \ 
r_\t{s}=\frac{E^{\t{s},\t{s}}_{0x}}{E^\t{i}_{0x}},\ 
r_\t{a}=\frac{E^{\t{s},\t{a}}_{0x}}{E^\t{i}_{0x}},
\label{r:decomp}\\
t &=& \frac{E^\t{t}_{0x}}{E^\t{i}_{0x}}=t_\t{s}-r_\t{a},\ \ 
t_\t{s}=1+r_\t{s},
\label{t:decomp}
\end{eqnarray}
where $r_\t{s}$ and $r_\t{a}$ represent reflection coefficients
for the symmetric and anti-symmetric components, respectively.
In (\ref{t:decomp}), $t_s$ denotes the transmission coefficient
in the absence of anti-symmetric scattering.

\section{Transmission phase for lossless, reflectionless metasurfaces}
\label{sec:lossless}

For lossless metasurfaces, power conservation requires
$|r|^2+|t|^2=1$. In terms of $t_\t{s}$ and $r_\t{a}$, this
condition translates into
\begin{equation}
|t_\t{s}|^2+|r_\t{a}|^2-\Re\{t_\t{s}+r_\t{a}\}=0.
\label{cond_lossless:tsra}
\end{equation}
Now, we require a reflectionless operation for maximum power
transmission $(r=0)$, or set $r_\t{a}=1-t_\t{s}$. Using this
condition in (\ref{cond_lossless:tsra}), we obtain
$|t_\t{s}|^2=\Re\{t_\t{s}\}$, which can be rewritten as
\begin{equation}
|t_\t{s}|=\cos\phi_{t_\t{s}},
\label{cond_lossless:tts}
\end{equation}
if we write $t_\t{s}=|t_\t{s}|e^{j\phi_{t_\t{s}}}$
(i.e, $\angle t_\t{s}=\phi_{t_\t{s}}$). Using (\ref{cond_lossless:tts}),
the total transmission coefficient is found to be
\begin{equation}
\begin{split}
t &= t_\t{s}-r_\t{a}=2t_\t{s}-1=2\Re\{t_\t{s}\}-1+j2\Im\{t_\t{s}\}\\
&= 2|t_\t{s}|\cos\phi_{t_\t{s}}-1+j2|t_\t{s}|\sin\phi_{t_\t{s}}\\
&= \cos2\phi_{t_\t{s}}+j\sin2\phi_{t_\t{s}}=e^{j2\phi_{t_\t{s}}}.
\end{split}
\label{t:lossless}
\end{equation}
Hence, we find
\begin{equation}
\angle t=2\phi_{t_\t{s}}.
\label{anglet_lossless_ts}
\end{equation}
Starting from $t=1+2r_\t{s}$ instead, a similar analysis finds
\begin{equation}
\angle t=2\phi_{r_\t{s}}+\pi,
\label{anglet_lossless_rs}
\end{equation}
where $\phi_{r_\t{s}}=\angle r_\t{s}$.
From (\ref{anglet_lossless_ts})--(\ref{anglet_lossless_rs}),
we find that
the transmission phase range is twice those of the transmission
and reflection coefficients of the symmetric component.
The magnitude of the transmission coefficient is unity due to
full transmission, as it should be for lossless, 
reflectionless metasurfaces.  

Let us assume no bianisotropy and a standard Lorentz frequency
dispersion for the lossless electric polarizability $\alpha_\ee$
for the horizontal electric dipole $p_x$.
The polarizability can be modeled as
\cite{bohren2004,asadchy_prx2015}
\begin{equation}
\alpha_\ee=\frac{A_\t{e}}{\omega_\t{e}^2-\omega^2},
\label{alphaee_model}
\end{equation}
where $A_\t{e}$ is the resonance strength coefficient related 
to the plasma frequency and $\omega_\t{e}$ is the electric resonance
frequency. The polarizability $\alpha_\ee$ relates the induced
dipole moment to the local excitation field at the dipole location.
By defining an effective polarizability $\hat{\alpha}_\ee$
for the dipole in an array environment, the induced dipole moment
can be written relative to the incident field as
$p_x=\hat{\alpha}^\ee E^\t{i}_{0x}$. It is known that the two
polarizabilities are related by \cite{asadchy_prx2015}
\begin{equation}
\frac{1}{\eta\hat{\alpha}_\ee}=\frac{1}{\eta\alpha_\ee}
+\frac{j\omega}{2S}\cos\theta^\t{i}
\label{alphaee_eff}
\end{equation}
in the TM polarization.
Using (\ref{def:Ess0x}), (\ref{alphaee_model}), and
(\ref{alphaee_eff}), the expression for $r_\t{s}$ is found to be
\begin{equation}
r_\t{s}=-j\frac{\omega\cos\theta^\t{i}}{2S}
\left(\frac{\omega_\t{e}^2-\omega^2}{\eta A_\t{e}}
+\frac{j\omega}{2S}\cos\theta^\t{i}
\right)^{-1}.
\end{equation}
At a design frequency $\omega$,
the phase of $r_\t{s}$ can be adjusted to any value in the range
$[-\pi,-\pi/2]$ or $[\pi/2,\pi]$
by adjusting $A_\t{e}$, $\omega_\t{e}$, and $S$ of a resonant meta-atom.
From (\ref{anglet_lossless_rs}), it can be seen
$\angle t$ has a full $2\pi$ range.

\begin{figure}[t]
 \centering
 \includegraphics{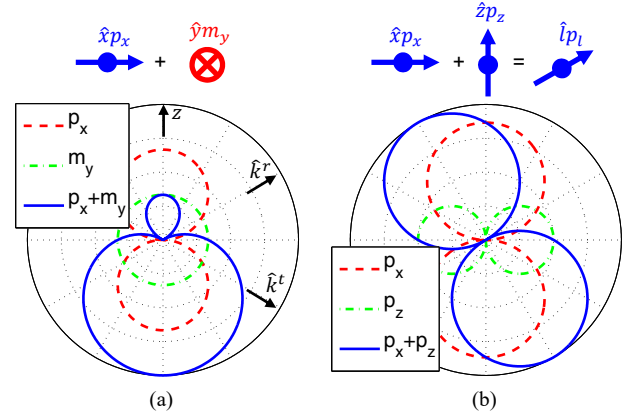}
 \caption{Meta-atom scattering magnitude
patterns for zero reflection at $\theta^\t{i}=60\degree$.
(a) A combination of $\xhat p_x$ and $\yhat m_y$.
(b) A combination of $\xhat p_x$ and $\zhat p_z$.
The total pattern as well as the patterns of elemental
dipoles are shown. The patterns are in linear scale with
an arbitrary unit. All dipoles are treated as point dipoles.}
 \label{fig:patterns_comparison}
\end{figure}

It is instructive to inspect the physical conditions for
zero reflection in terms of dipole moments.
For a combination of $\xhat p_x$ and $\yhat m_y$,
setting $r=r_\t{s}+r_\t{a}=0$ using (\ref{def:Ess0x}) and
(\ref{def:Esa0x_my}) gives
\begin{equation}
m_y+\eta p_x\cos\theta^\t{i}=0.
\label{cond:norefl_pxmy}
\end{equation}
{\color{red}In \cite{paniagua-dominguez_natcommun2016},
extended relations of (\ref{cond:norefl_pxmy})
between the electric and magnetic dipole amplitudes for
a generalized Brewster effect in dielectric metasurfaces
were derived for any given incidence angle, frequency,
and polarization.}
For normal incidence $(\theta^\t{i}=0)$,
{\color{red}(\ref{cond:norefl_pxmy})} corresponds to
a zero backscattering condition (in the $+z$-axis direction).
This condition for an orthogonal set of electric and magnetic
dipoles has also been utilized in antenna 
designs~\cite{green_ieeejap1966,jin_ieeejawpl2010}.
In the case of planar array or metasurface designs, zero
specular reflection is of interest and (\ref{cond:norefl_pxmy})
is the necessary condition at any incidence angle.
In \cite{radi_prappl2015}, the same balance condition
was the principle behind thin Huygens sheet absorbers under
a normal incidence. As an example, the element (meta-atom)
scattering pattern of a balanced electric-magnetic dipole
pair for a 60\textdegree{} incidence case is shown
in Fig.~\ref{fig:patterns_comparison}(a). Together with
the total pattern, the individual scattering patterns of
$\xhat p_x$ and $\yhat m_y$ are also shown. In the specular
reflection $(\khat^\t{r})$
direction, the two individual patterns destructively
interfere to create a null, while they interfere constructively
in the transmission $(\khat^\t{t})$ direction.
The phases of the two moments $p_x$ and $m_y$ may
be varied under the condition
(\ref{cond:norefl_pxmy}) to tune the transmission phase.

For a combination of $\xhat p_x$ and $\zhat p_z$, the zero
reflection condition translates to
\begin{equation}
p_x\cos\theta^\t{i}=p_z\sin\theta^\t{i}.
\label{cond:norefl_pxpz}
\end{equation}
Hence, the two orthogonal induced dipole components should be
in phase and their magnitudes should scale properly as a function
of $\theta^\t{i}$.
The individual and total scattering patterns for a combination
of two electric dipoles satisfying (\ref{cond:norefl_pxpz})
are shown in Fig.~\ref{fig:patterns_comparison}(b) for the
same 60\textdegree{} oblique illumination. A destructive
interference creates a scattering null in the $\khat^\t{r}$
direction. As indicated in Fig.~\ref{fig:patterns_comparison}(b),
this combination of two orthogonal dipoles corresponds to a single
tilted electric dipole. Hence, there exists an interesting
realization strategy in microwaves, where the direction of
a low-loss conduction current can be accurately controlled
by shaping thin conductors in meta-atoms.
{\color{red}Using a tilted straight dipole meta-atom
to realize $\hat{l}p_l$ in Fig.~\ref{fig:patterns_comparison}(b),
(\ref{cond:norefl_pxpz}) will be satisfied for any value of
$p_l$. The tangential and normal dipole moments will follow exactly
the same frequency dispersion. The result is dispersionless
zero reflection at the specific oblique incidence angle that
matches the dipole tilt angle in the opposite direction. For
lossless meta-atoms, this translates into dispersionless full
transmission.}
When volumetric
electric polarization currents are utilized, $\xhat p_x$ and 
$\zhat p_z$ should be jointly designed to satisfy
the proper magnitude and phase relations, i.e.,
(\ref{cond:norefl_pxpz}). 
A microwave metasurface design based on the former strategy
and an optical design based on the latter approach are
presented in Sec.~\ref{sec:examples}.

It is stressed that using tangential and normal electric dipoles
for creating reflectionless metasurfaces is limited to oblique
illuminations in the TM polarization.
At a normal incidence, the reflection
direction is also normal to the surface. Even if
the normal electric dipole, $\zhat p_z$, is induced by the
incident wave, it cannot scatter in the reflection and transmission
directions, which are the axial directions of $\zhat p_z$.
In comparison, the combination of tangential electric
and magnetic dipoles is available for any angle of incidence.

Since either a tangential magnetic dipole or a normal electric
dipole can contribute to creating zero reflection 
together with a given
tangential electric dipole, we can find
a relation between $m_y$ and $p_z$ for contributing the same
effect. By equating the values of $r_\t{a}$ due to the two
dipoles from (\ref{def:Esa0x_pz}) and (\ref{def:Esa0x_my}),
we find $m_y=-\eta p_z\sin\theta^\t{i}$, which can be written
in the form
\begin{equation}
\yhat M_y=-\nhat\times\frac{\bk_t}{\omega\epsilon_0}J_z;\ \ 
\nhat=\zhat,\ \bk_t=\xhat k\sin\theta^\t{i}
\label{equiv:my_pz}
\end{equation}
in terms of surface-averaged polarizations.
Here, $\epsilon_0$ is the free-space permittivity and
$\nhat$, $\bk_t$ represent the unit surface normal and the
tangential wave vector, respectively.
Identified in \cite{albooyeh_prb2017}, (\ref{equiv:my_pz}) represents
an equivalence between a tangential magnetic
polarization and a spatially-varying normal electric polarization.

\section{Transmission magnitude and phase for lossy, reflectionless
metasurfaces}
\label{sec:lossy}

Absorption in lossy constituent materials
is inevitable in practical realizations, resulting in reduction
of the transmission power efficiency from 100\%.
On the other hand, zero reflection and zero transmission
are desired for absorber applications. In this Section,
we investigate if there is a limitation in designing the
transmission phase when absorption is present and how the phase
may be controlled.

Let the absorptance in the scattering described in
Fig.~\ref{fig:tm_refl_trans} be denoted by $A$ in the range
$0<A<1$. It is defined as the absorbed power within a unit cell normalized
by the incident power on the unit-cell area.
Then, the power relation dictates $|r|^2+|t|^2=1-A$.
Using the decompositions of $r$ and $t$ into the
symmetric and anti-symmetric
scattering components (\ref{r:decomp})--(\ref{t:decomp}), this
power conservation relation can be written in terms of $t_\t{s}$ and
$r_\t{a}$ as
\begin{equation}
\Re\{t_\t{s}+r_\t{a}\}-\left(|t_\t{s}|^2+|r_\t{a}|^2\right)=\frac{A}{2}.
\label{power_conserv_lossy}
\end{equation}
For both high-transmission and absorber applications, zero
reflection is desirable. Using $r_\t{a}=1-t_\t{s}$ in
(\ref{power_conserv_lossy}) gives
\begin{equation}
|t_\t{s}|^2-\Re\{t_\t{s}\}+\frac{A}{4}=0.
\label{eq:ts_lossy}
\end{equation}
The solution for $|t_\t{s}|$ is readily found to be
\begin{equation}
|t_\t{s}|=\frac{\cos\phi_{t_\t{s}}\pm\sqrt{\cos^2\phi_{t_\t{s}}-A}}{2}.
\label{sol:tsmag_lossy}
\end{equation}
We find that the passivity and power conservation principles
do not require a particular one of the two 
branches to be taken for  the square root function in
(\ref{sol:tsmag_lossy}). Both values of
$|t_\t{s}|$ are valid solutions. For real-valued
solutions to exist for $|t_\t{s}|$, it is required that
$\cos^2\phi_{t_\t{s}}-A\geq 0$. Hence, we find that $t_\t{s}$ has
its phase limited to the range
\begin{equation}
-\cos^{-1}\sqrt{A}\leq\phi_{t_\t{s}}\leq\cos^{-1}\sqrt{A}.
\label{range_phits_lossy}
\end{equation}
Using (\ref{sol:tsmag_lossy}), the transmission coefficient
is expressed as
\begin{equation}
t=e^{j\phi_{t_\t{s}}}\left(\pm\sqrt{\cos^2\phi_{t_\t{s}}-A}+j\sin\phi_{t_\t{s}}\right).
\label{t:lossy}
\end{equation}
If we carry out a similar analysis in terms of $r_\t{s}$ instead
of $t_\t{s}$, it is found that the range of $\phi_{r_\t{s}}$ is limited
to
\begin{equation}
\cos^{-1}\left(-\sqrt{A}\right)\leq|\phi_{r_\t{s}}|\leq\pi.
\label{range_phirs_lossy}
\end{equation}
The transmission coefficient has an alternative expression
given by
\begin{equation}
t=e^{j\phi_{r_\t{s}}}\left(\pm\sqrt{\cos^2\phi_{r_\t{s}}-A}-j\sin\phi_{r_\t{s}}\right).
\label{t_alt:lossy}
\end{equation}
By allowing both branches in (\ref{t:lossy}) and
(\ref{t_alt:lossy}), it is easy to verify that the range of
transmission phase is $2\pi$ regardless of the value of $A$.
Hence, we conclude that
the presence of absorption does not fundamentally
require the transmission phase range to be reduced from
$2\pi$ associated with the lossless case.

Using a lossy meta-atom model, we inspect the range of a realizable
transmission phase and develop a design approach for achieving
a particular combination of $|t|$ and $\angle t$. A
lossy meta-atom having a Lorentz-type resonant response
can be modeled using a polarizability function given by
\begin{equation}
\alpha_\ee=\frac{A_\t{e}}
{\omega_\t{e}^2-\omega^2+j\gamma_\t{e}\omega},
\label{alphaee_model_lossy}
\end{equation}
where $\gamma_\t{e}$ is the electric loss factor or collision
frequency. Using (\ref{def:Ess0x}), (\ref{alphaee_eff}), and
(\ref{alphaee_model_lossy}), the transmission coefficient,
written as $t=1+2r_\t{s}$ under a zero reflection condition,
is expressed as
\begin{equation}
t=\frac{\frac{\omega_\t{e}^2-\omega^2}{\eta A_\t{e}}+
j\omega\left(\frac{\gamma_\t{e}}
{\eta A_\t{e}}-\frac{\cos\theta^\t{i}}{2S}\right)}
{\frac{\omega_\t{e}^2-\omega^2}{\eta A_\t{e}}+
j\omega\left(\frac{\gamma_\t{e}}
{\eta A_\t{e}}+\frac{\cos\theta^\t{i}}{2S}\right)}.
\label{t:lossy_lorentz}
\end{equation}
For notational simplicity, let us introduce symbols
$u=\gamma_\t{e}\omega/\eta A_\t{e}$,
$v=\omega\cos\theta^\t{i}/2S$, 
and $w=(\omega_\t{e}^2-\omega^2)/\eta A_\t{e}$.
It is noted that $u$ and $v$ are positive quantities, but
$w$ can take any real value. Enforcing an absorptance value of
$A=1-|t|^2$ relates $u$, $v$, $w$, and $A$. The resulting value of $w$
can be written in terms of the remaining quantities as
\begin{equation}
w=\pm\sqrt{2\left(\frac{2}{A}-1\right)uv-u^2-v^2}.
\label{w:lossy}
\end{equation}
For $w$ to be real-valued, the quantity under the square root
should be non-negative. This defines the region of a valid point
$(u,v)$ in the $uv$-plane. It is found that the ratio $u/v$ is
bound between two constants defined by $A$, i.e.,
\begin{multline}
2\left[\frac{1}{A}-\sqrt{\frac{1}{A}\left(\frac{1}{A}
-1\right)}\right]-1<\frac{u}{v}\\
<2\left[\frac{1}{A}+\sqrt{\frac{1}{A}\left(\frac{1}{A}
-1\right)}\right]-1.
\label{uoverv_range}
\end{multline}
Let us denote the lower and upper limits in
(\ref{uoverv_range}) by $s_{0}$ and $s_{1}$, respectively.
It is noted that
\begin{equation}
0<s_0<1,\ s_1>1,\ \text{and}\ s_0s_1=1,
\label{properties:s0s1}
\end{equation}
for all possible values of $A$.
In addition, let us introduce a slope function $s$ as a function 
of a parameter $q$ via
\begin{equation}
s(q)=s_{0}+(s_{1}-s_{0})q,\ \ 0<q<1,
\label{def:sr}
\end{equation}
so that $u=s(q)v$.

\begin{figure}[t]
 \centering
 \includegraphics*{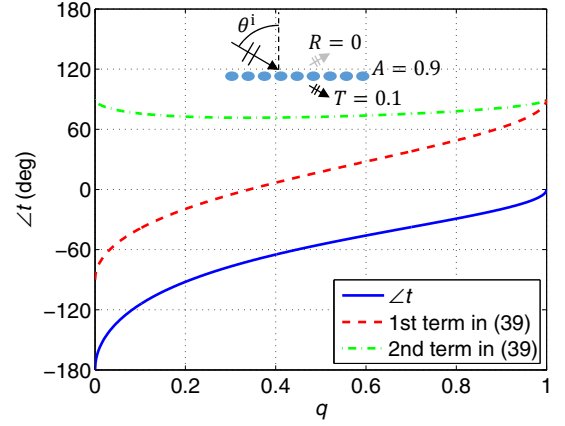}
 \caption{Transmission phase design for a lossy,
reflectionless metasurface with an
absorptance of $A=0.9$. The values of a reflectance $R=0$
and a transmittance $T=0.1$ are also indicated in the inset.
The range of $u/v$ for this example is equal to
$0.520<u/v<1.925$.}
 \label{fig:tphase_range}
\end{figure}

For the time being, let us assume that $w>0$ in (\ref{w:lossy}).
From (\ref{t:lossy_lorentz}), the transmission phase is expressed as
\begin{equation}
\angle t=\tan^{-1}\frac{u/v-1}{|w|/v}
-\tan^{-1}\frac{u/v+1}{|w|/v}.
\label{anglet:lossy_lorentz}
\end{equation}
In (\ref{anglet:lossy_lorentz}), the quantity $|w|/v$ is positive
and $|w|/v\to 0^+$ as $s\to s_0$, $s_1$. Now, we inspect the
numerators in the argument of the arctangent functions in
(\ref{anglet:lossy_lorentz}).
For the second arctangent, it is
seen that $u/v+1$ remains positive in $0<q<1$
(i.e., its entire range) for all possible values of
$A$. The numerator in the first arctangent
function has a range
\begin{equation}
s_0-1<\frac{u}{v}-1<s_1-1.
\label{atan1_num}
\end{equation}
From (\ref{properties:s0s1}),
we note that the lower limit in (\ref{atan1_num})
is a negative quantity,
while the upper limit is a positive one, for all possible
$A$. Hence, the first term in
(\ref{anglet:lossy_lorentz}) changes from $-\pi/2$ to $\pi/2$
as $q$ is increased from 0 to 1. The second term in
(\ref{anglet:lossy_lorentz}) stays in the range between
0 and $\pi/2$, but it reaches $\pi/2$ at $q=0$, $1$. Hence,
we expect a range of $\pi$ for the transmission phase, in
$-\pi<\angle t<0$. For an example case of $A=0.9$,
Fig.~\ref{fig:tphase_range} shows
the transmission phase with respect to $q$. It can be
seen that there exists a unique value of $q$ for achieving
a desired value of $\angle t$.

For synthesizing a transmission phase in $-\pi<\angle t<0$,
a meta-atom design strategy can be described as follows.
For a given absorptance $A$, the transmission magnitude is
fixed at $|t|=\sqrt{1-A}$ for a reflectionless response.
The phase angle $\angle t$ spans $[-\pi,0]$ as a function
of $u/v$ in $s_0<s<s_1$ $(0<q<1)$. 
Equation~(\ref{anglet:lossy_lorentz}) can be solved for
the value of $q$ that achieves a desired transmission phase
and the corresponding value of $u/v$ follows.
The associated value of $w/v$ is obtained from (\ref{w:lossy}).
At the design frequency $\omega$ and the incidence angle
$\theta^\t{i}$, a meta-atom is designed by determining
a combination
of values for $A_\t{e}$, $\omega_\t{e}$, $S$, and $\gamma_\t{e}$ for
giving the determined ratios $u/v$ and $w/v$.

A transmission phase in the range $0<\angle t<\pi$ can be
designed using the negative branch for $w$ in (\ref{w:lossy}).
For $w<0$, we note from (\ref{t:lossy_lorentz}) that
\begin{equation}
\angle t=-\left(\tan^{-1}\frac{u/v-1}{|w|/v}
-\tan^{-1}\frac{u/v+1}{|w|/v}\right),
\label{angletpos:lossy_lorentz}
\end{equation}
which is an opposite number for (\ref{anglet:lossy_lorentz}).
Therefore, to design a value of $\angle t$ in $[0,\pi]$,
meta-atom parameters can be first determined for achieving a phase of
$-\angle t$, following the approach described above for
realizing a transmission phase in $[-\pi,0]$. Then,
the sign of $w$ needs to be changed, which can be achieved
by setting $\omega_\t{e}<\omega$, i.e., by choosing the resonance
frequency of the lossy meta-atom lower than the design frequency.

Combining the two separate cases of the desired transmission phase
being in $[-\pi,0]$ or $[0,\pi]$, we conclude that 
a reflectionless metasurface can be designed using 
lossy meta-atoms
following the Lorentz dispersion model to achieve
a transmission phase in a full $2\pi$ range at any level of
absorption at any oblique angle of incidence.

\section{Numerical examples}
\label{sec:examples}

In this section, we will  present different alternatives for 
Huygens' metasurfaces where the tangential and normal 
electric dipole moments are carefully engineered for 
satisfying the reflectionless condition
while providing a complete transmission phase coverage.
We will explore the possibilities for the design of 
metasurfaces at microwave and optical frequencies for 
lossless and lossy scenarios.

\subsection{Reactively-loaded tilted thin 
conductor strip dipole array}
\label{sec:tilted_pec_array}

\begin{figure}[t]
 \centering
 \includegraphics*{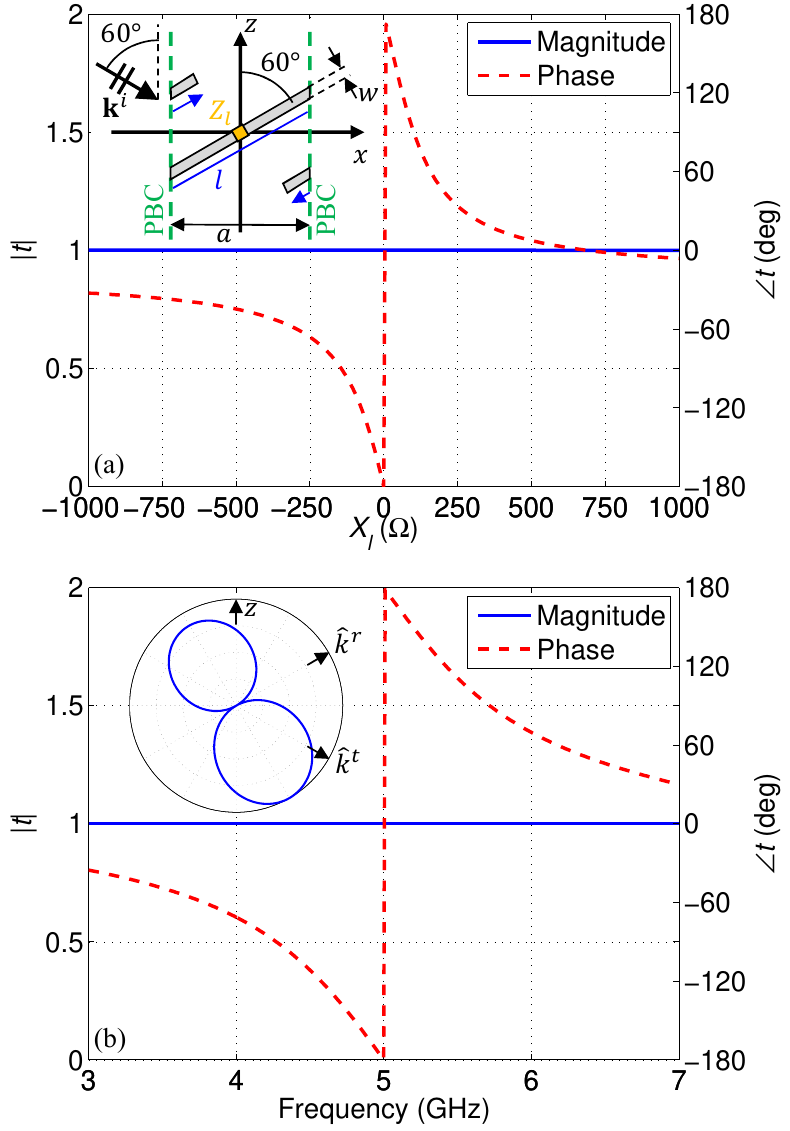}
 \caption{Plane wave scattering by a planar array of tilted
conductor strip dipoles. The meta-atom comprises a conducting
strip dipole of dimensions $l\times w$. 
The plane of the strip is the $xz$-plane
and the strip surface is assumed to be perfectly conducting.
A point load with a reactive impedance $Z_l=jX_l$ is connected
at the dipole middle point.
(a) The transmission magnitude and phase with respect to the
load reactance $X_l$ at 5~GHz.
The inset shows a side view of the
unit cell together with the illuminating plane wave.
(b) The transmission coefficient as a function of frequency
when the load port is shorted $(X_l=0)$.
The unit-cell dimension is $a=20~\text{mm}$ in both $x$-
and $y$-axis directions.
The length and width of the strip dipole are
$l=29.9~\text{mm}$ and $w=0.6~\text{mm}$.
{\color{red}The unit-cell scattering pattern at 5~GHz
is shown in the inset.}}
 \label{fig:tilted_dipole_tmagphase}
\end{figure}

At microwave frequencies, the direction of induced electric dipole
moments can be easily controlled using thin conductor wires and
traces. For supporting both tangential and normal electric
dipoles, a straight thin conductor dipole can be tilted to
%satisfy
%the zero reflection condition (\ref{cond:norefl_pxpz}).
{\color{red}produce zero reflection. Here, it is noted that
(\ref{cond:norefl_pxpz}) is the zero reflection condition for
point dipoles. For meta-atoms of practical dimensions,
zero reflection corresponds to a scattering pattern null in the
specular reflection direction. For a straight conductor dipole
of any length, aligning the dipole axis in the specular reflection
direction guarantees $r=0$.}
The transmission characteristics of a tilted straight dipole
array are analyzed in Fig.~\ref{fig:tilted_dipole_tmagphase}.
For a TM-polarized plane-wave illumination at an incidence angle
of 60\textdegree{}, a perfect electric conductor (PEC) strip
dipole tilted at the same 60\textdegree{} angle in the opposite
direction constitutes the meta-atom, as shown in the inset of
Fig.~\ref{fig:tilted_dipole_tmagphase}(a). At the middle point of the
dipole, a lumped load with an impedance $Z_l$ is connected.
{\color{red}While providing tunability for $t$,
such a load does not affect the reflectionless property.}
At a design
frequency of 5~GHz, the unit-cell dimensions are chosen to be
$a=b=20~\text{mm}$, so that there are no higher-order propagating
Floquet modes. For an
electrically thin width of $w=0.6~\text{mm}$, the length
of the strip was adjusted to $l=29.9~\text{mm}$
such that the transmission
phase for the unloaded case ($Z_l=0$) is equal to $\pi$.
The dipole meta-atom extends slightly into neighboring cells.
Using a phase-shift periodic boundary condition (PBC) on the
four vertical walls of the unit cell, the scattering
characteristics of an infinitely large planar array were
simulated using FEKO 2017 by Altair.

At the design frequency,
the magnitude and phase of $t$ are plotted in
Fig.~\ref{fig:tilted_dipole_tmagphase}(a) with respect to the
reactance $X_l$ of the reactive load impedance $Z_l=jX_l$.
Since the meta-atom is lossless and the dipole does not
reflect, the transmission magnitude is constant at unity.
By loading the dipole with different reactance $X_l$,
the transmission phase can be adjusted to any value in
the range $-\pi<\angle t\leq\pi$. Not the entire $2\pi$
phase range is visible in 
Fig.~\ref{fig:tilted_dipole_tmagphase}(a).
Simulations with large loading reactances
show that the transmission
phase approaches a single value of
$\angle t=-19.6\degree$ from different directions as
$X_l\to\pm\infty$. Considering that the dipole meta-atom
is not electrically short, large loading reactances will not
make it effectively non-existent. Instead, the meta atom will
appear as two narrowly separated collinear dipoles of a length
$l/2$ each.

For an unloaded dipole $(X_l=0)$, 
Fig.~\ref{fig:tilted_dipole_tmagphase}(b) plots the transmission
coefficient with respect to frequency. A standard Lorentz-type
resonant frequency response is observed for the phase.
{\color{red}In the inset, the normalized unit-cell scattering
pattern at 5~GHz in the $xz$-plane is shown as a polar plot.
The length of the dipole meta-atom of nearly a half wavelength
makes the pattern deviate noticeably from that of a point
dipole shown in Fig.~\ref{fig:patterns_comparison}(b).
Still, a scattering null is synthesized in the reflection direction
due to the tilt.}  
In this design, 
the bandwidth of full power transmission is in principle 
infinite. This is due to satisfaction of the no reflection
condition of a geometrical origin. For a tilted straight dipole,
the normal and tangential surface-averaged
electric polarizations,
$J_x=j\omega p_x/S$, $J_z=j\omega p_z/S$, have the exact same
frequency dispersion. If zero reflection is achieved via
a balance of electric and equivalent magnetic polarizations
as is done in Huygens' metasurfaces, the high-transmission
frequency bandwidth is not expected to be wide.

\subsection{Impedance-loaded tilted strip dipole array}
\label{sec:tilted_array_lossy}

\begin{figure}[t]
 \centering
 \includegraphics*{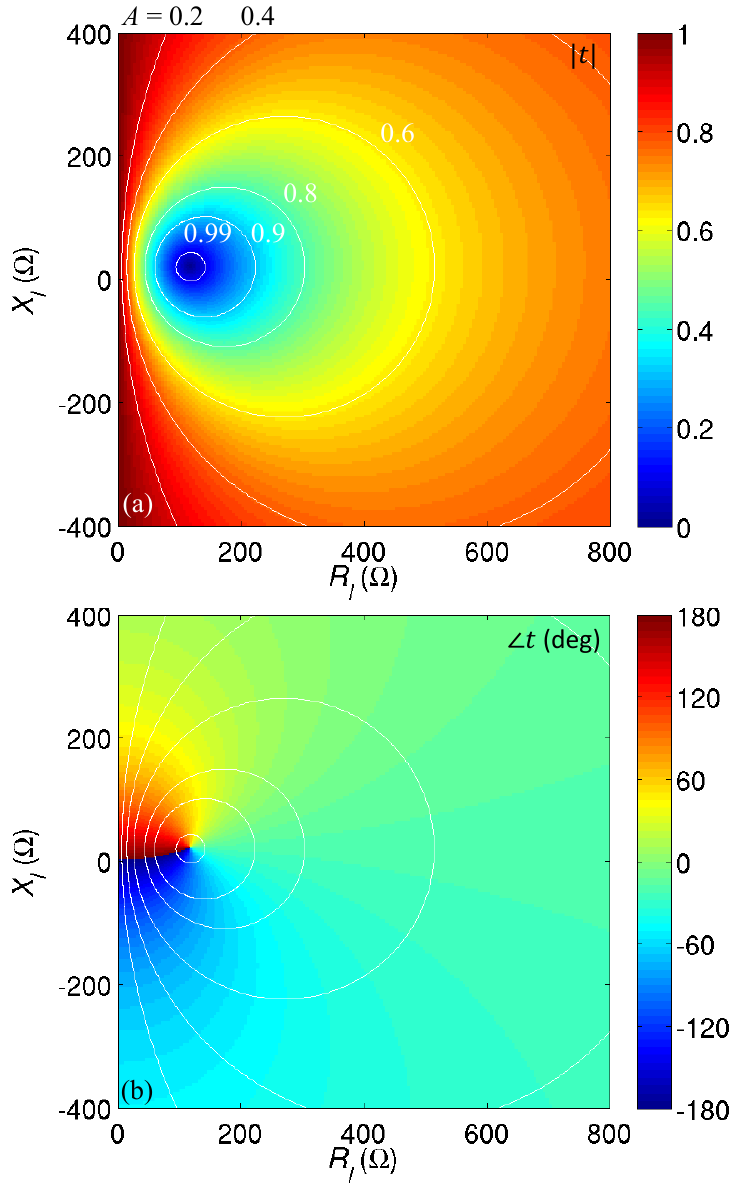}
 \caption{The transmission coefficient with respect to the
lumped load resistance $R_l$ and reactance $X_l$ values.
(a) The magnitude. (b) The phase angle in degrees.
The white contour lines correspond to different absorptance
values of $A=\{0.2, 0.4, 0.6, 0.8, 0.9, 0.99\}$.
The $A=0$ contour is the $R_l=0$ axis.}
 \label{fig:tilted_strip_rxload}
\end{figure}

To the tilted conductor dipole array of 
Sec.~\ref{sec:tilted_pec_array}, loss can be introduced to
make the metasurface absorptive. For this purpose, a resistive
component can be incorporated into the load impedance
$Z_l$ in Fig.~\ref{fig:tilted_dipole_tmagphase}(a).
At the same time, adjusting the reactance part is expected to
give a capability to tune the transmission phase. To the
tilted PEC strip dipole array considered in
Sec.~\ref{sec:tilted_pec_array}, a complex load with an impedance
$Z_l=R_l+jX_l$ is attached to the center point of the
meta-atom. In the range $0\leq R_l\leq 800~\Omega$,
$-400~\Omega\leq X_l\leq 400~\Omega$, the reflection and transmission
coefficients were simulated using FEKO at 5~GHz.
The simulated 
reflection coefficient is zero. The transmission
coefficient is shown in Fig.~\ref{fig:tilted_strip_rxload}.
The magnitude plot in
Fig.~\ref{fig:tilted_strip_rxload}(a) demonstrates that
the entire range of $|t|$ between zero and unity
is available for synthesis. 
For visualization, a few contours for constant-$|t|$ 
(in terms of the absorptance $A$) values are also plotted. The same
contours are reproduced in Fig.~\ref{fig:tilted_strip_rxload}(b),
where the transmission phase is plotted. It is clear that
for any given value of $A$, an arbitrary phase in a complete
$2\pi$ range can be achieved in principle by selecting 
an appropriate value for $Z_l$. In practice, extreme values
of $R_l$, $X_l$ required for some combinations of
$|t|$ and $\angle t$ may be difficult to realize and
consequently limit the synthesis range.

\begin{figure}[t]
 \centering
 \includegraphics*{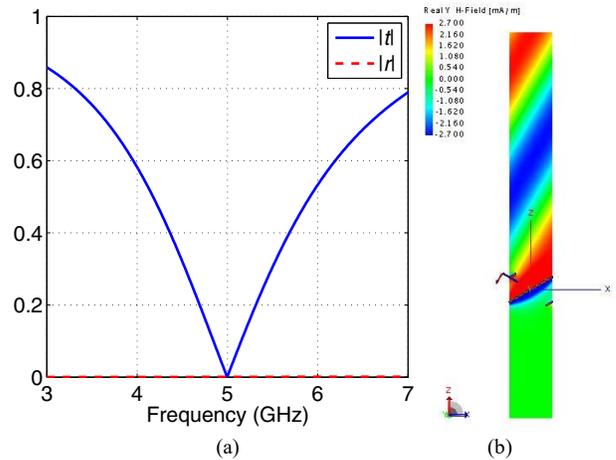}
 \caption{Characteristics of the center-loaded strip dipole
array with the load impedance chosen for perfect absorption at 5~GHz.
(a) The magnitude of the reflection and transmission
coefficients with respect to frequency.
(b) A snapshot of the total $y$-directed magnetic field,
$H_y$.}
 \label{fig:tilted_strip_absorber}
\end{figure}

In Fig.~\ref{fig:tilted_strip_rxload}(a), an extreme case
of full absorption is observed with a selection of load impedance
$Z_l=115.9+j20.6~\Omega$. If the received power is guided
to a receiving circuitry rather than dissipated as heat,
the design corresponds to a planar receiving array with a 100\% receiving
efficiency. The frequency responses of $|r|$ and $|t|$ are
shown in Fig.~\ref{fig:tilted_strip_absorber}(a).
The metasurface does not transmit nor reflect at the design
frequency of 5~GHz. The bandwidth is reflection zero is extremely
wide. At 5~GHz, a snapshot at time $t=0$
of the $y$-component of the total magnetic field 
in the $xz$-plane is plotted over the unit-cell dimension of
$-a/2<x<a/2$ in Fig.~\ref{fig:tilted_strip_absorber}(b). 
Only the fields
associated with the incident wave is visible above the
metasurface and zero field penetrates behind the dipole array.

In this design, absorption occurs at the load connected
to the otherwise lossless meta-atom. Instead, it is possible to design
an absorber based on a distributed loss mechanism. One approach
will be realizing the dipole meta-atom using lossy conductive
material such as conductive ink films,
modeled as a resistive impedance sheet~\cite{wang_ieeejmtt2017}.
Such a lossy strip array variant was designed for perfect
absorption at 5~GHz and numerically analyzed.
Its simulated scattering characteristics exhibit
the same qualitative behavior presented in
Fig.~\ref{fig:tilted_strip_absorber} (not shown).

\subsection{Array of dielectric bars of rectangular cross section}
\label{sec:dielectric_cylinders}

\begin{figure}[t]
	\centering
%	\subfigure[]{\includegraphics[width=0.45\textwidth]{Reflection_separated_peaks}\label{fig:dielectric_bars_reflection_non_optimized_a}}\\
%	\subfigure[]{\includegraphics[height=0.18\textwidth]{Magnetic_111THz}\label{fig:dielectric_bars_reflection_non_optimized_b}}
%	\subfigure[]{\includegraphics[height=0.18\textwidth]{Electric_tangential_157THz}\label{fig:dielectric_bars_reflection_non_optimized_c}}
%	\subfigure[]{\includegraphics[height=0.18\textwidth]{Electric_vertical_177THz}\label{fig:dielectric_bars_reflection_non_optimized_d}}
 \includegraphics{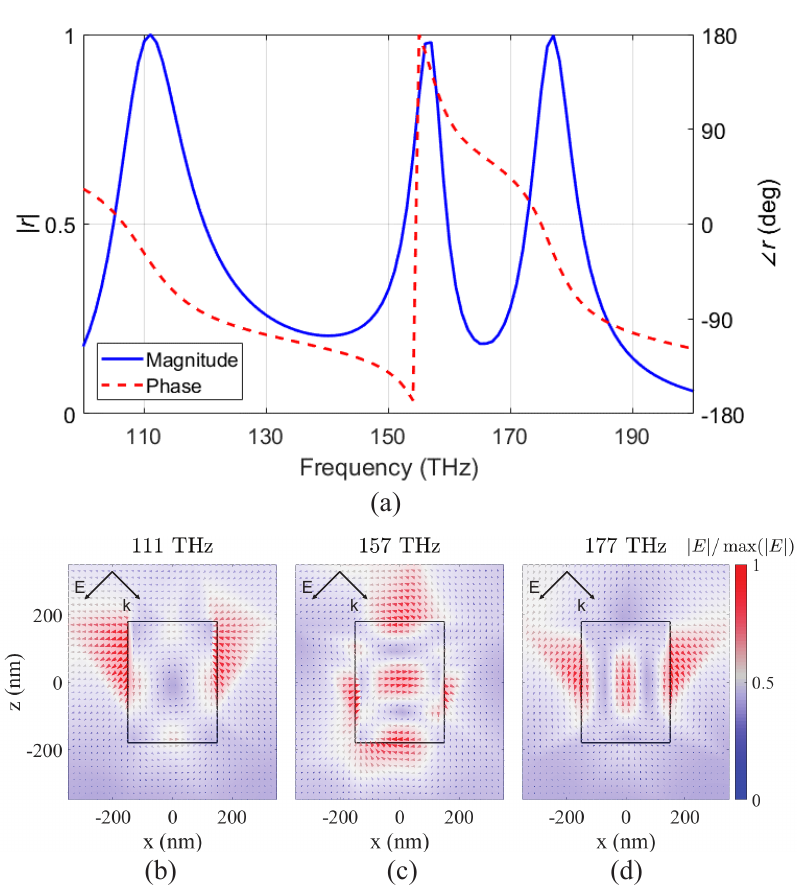}
 \caption{Scattering by an array of dielectric bars illuminated
obliquely at an incidence angle of
$\theta^{\rm{i}}=45\degree$. The refractive index 
of the dielectric material is $n_{\rm d}=6$, the periodicity 
is $a=730$~nm, the width is $w=300$~nm and the height is $h=360$~nm.
(a) Reflection spectrum.
(b-d) Field distributions at the three resonant frequencies: 
(b) A tangential magnetic dipole;
(c) A tangential electric dipole; 
(d) A vertical electric dipole.}
 \label{fig:dielectric_bars_reflection_non_optimized}
\end{figure}

\begin{figure}[t]
	\centering
%	\subfigure[]{\includegraphics[width=0.23\textwidth]{scheme}}
%	\subfigure[]{\includegraphics[width=0.23\textwidth]{Transmission_shape_parameters}}\\
%	\subfigure[]{\includegraphics[width=0.23\textwidth]{Tilted_dipole_optimized}}
%	\subfigure[]{\includegraphics[width=0.23\textwidth]{tilted_dipole_concept}}
 \includegraphics{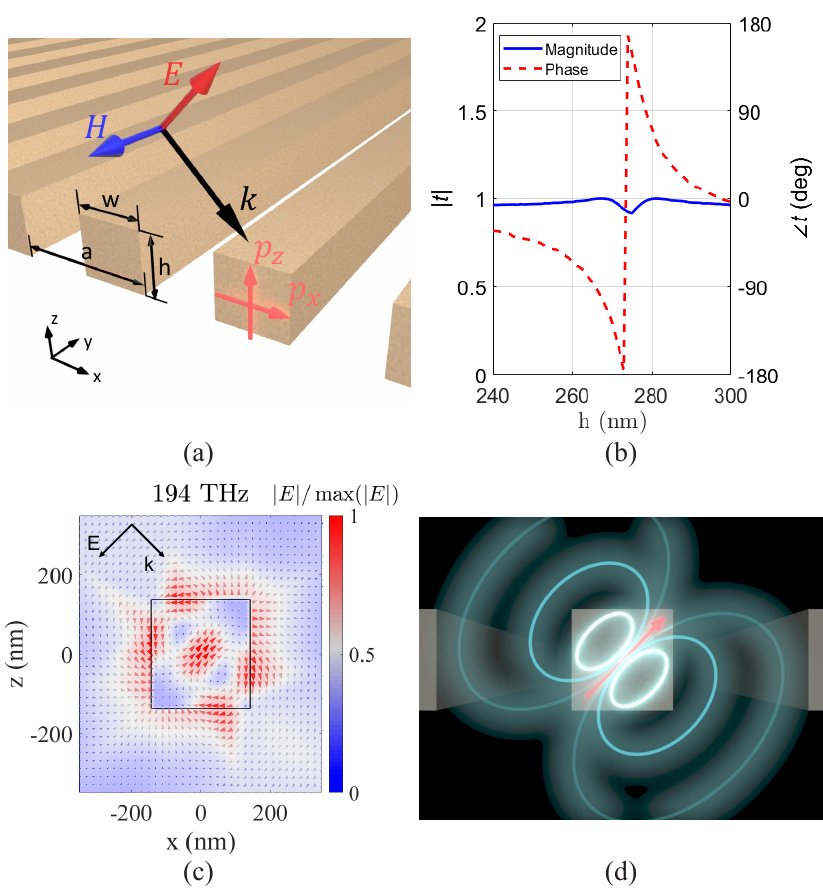}
	\caption{A high-transmission all-dielectric array
utilizing induced electric dipoles.
(a) Schematic of the studied structure:
an array of dielectric bars with a refractive index 
$n_{\rm d}=6$ illuminated at an angle of $\theta^{\rm i}=45\degree$.
(b) Transmission amplitude and phase of arrays of bars at 
a frequency of 194~THz with different shape parameters. 
The periodicity of the bars is kept constant, $a=730$~nm, 
and the width of the bars is $w=h+10~\text{nm}$.
(c) An electric field distribution at 194~THz for $h=275$~nm.
(d) A conceptual illustration of the induced dipole moment.}
	\label{fig:dielectric_bars_transmission_optimized}
\end{figure}

Here we give an example of a high-transmission array offering full control of the transmission phase in the optical domain. In order to ensure small dissipation losses, we use an all-dielectric metasurface formed by an array of parallel bars made of a lossless high refractive index dielectric ($n_{\rm d}=6$)
with the axis of the bars oriented along 
the $y$-direction. Due to the 2-D nature of the problem, 
the $y$-periodicity is infinite so we ensure the absence 
of higher order modes propagating in this direction.
In conventional realizations of high-transmission all-dielectric metasurfaces, the Huygens regime is realized by exciting both electric and magnetic moments at the operational frequency. Here, we illuminate the array by an obliquely propagating plane wave to ensure excitation of both tangential and normal electric polarizations and utilize the theory in this paper to realize 
a full transmission with a complete phase control. 
Following the previous example, we consider a TM-polarized incident plane wave [see Fig.~\ref{fig:dielectric_bars_transmission_optimized}(a)] and we define the incidence angle 
to be 45\textdegree.

The reflection spectrum is presented in Fig.~\ref{fig:dielectric_bars_reflection_non_optimized}(a) when the periodicity is $a=730$~nm, the width is $w=300$~nm and the height is $h=360$~nm. It exhibits three reflection maxima which correspond to three dipole-mode resonances. 
The first reflection peak, at 111~THz, corresponds to a
tangentially oriented magnetic dipole. Figure~\ref{fig:dielectric_bars_reflection_non_optimized}(b) shows the electric field vector in the $xz$-plane, where we can see that the electric field circulates inside the bars generating 
an out-of-plane magnetic dipole moment. 
At 157~THz, the reflection peak is caused by the resonance of 
a
tangential electric dipole, as we can see in Fig.~\ref{fig:dielectric_bars_reflection_non_optimized}(c). The third resonance peak, at 177~THz, corresponds to the resonance of a
normal electric dipole [see Fig.~\ref{fig:dielectric_bars_reflection_non_optimized}(d)]. Adjusting  the shape parameters of the bars, the magnitude and phase of the induced dipoles can be tuned to fulfill the conditions in (\ref{cond:norefl_pxpz}).

Finally, a combination of~$\hat{x}p_{x}$ and~$\hat{z}p_{z}$ can be tuned to transmit the
incident light with required phase variations within the $2\pi$ range. We consider a periodic array as shown in Fig.~\ref{fig:dielectric_bars_transmission_optimized}(a) with a period $a=730$~nm. The bars have a rectangular cross section with a height $h$ and a width $w=h+10$~nm. Figure~\ref{fig:dielectric_bars_transmission_optimized}(b) shows the transmission magnitude and phase for arrays of bars of different sizes, as a function of $h$. We see that the transmission phase indeed varies within the full $2\pi$ range while the transmission magnitude remains close to unity.  Figures~\ref{fig:dielectric_bars_transmission_optimized}(c,d) show the calculated field distribution and a conceptual illustration of an excited tilted electric dipole moment for $h=275$~nm. The simulations were performed using ANSYS~HFSS, setting PBCs on the vertical walls of the unit cell.

{\color{red}For absorptive applications, an array of dielectric
bars made from a suitable lossy material is a straightforward
extension of the high-transmission array in
Fig.~\ref{fig:dielectric_bars_transmission_optimized}.
An optical analogue of the array of
tilted dipoles in Sec.~\ref{sec:tilted_array_lossy}
is another possibility.
Using the approach of lossy surface impedance realization in
\cite{monti_oplett2016} exploiting the surface dispersion of
nanoparticles,
it can be envisaged that metal nanoparticles are deposited
on some dielectric support having periodic slanted walls.
The tilt angle of the dielectric support can enforce
(\ref{cond:norefl_pxpz}). The surface resistance may be
designed to achieve various levels of absorption, including
complete absorption. This approach needs further investigation.}

\section{Conclusion}
\label{sec:conclusion}

For polarization-preserving transmissive metasurface,
it has been shown that the origin of a full transmission 
capability together with a complete $2\pi$ phase coverage is  
synthesis of an asymmetric scattering pattern with respect to
the metasurface plane. In Huygens' meta atoms, a tangential
magnetic dipole is responsible for creating anti-symmetric
scattering. An alternative meta-atom arrangement is available
for full transmission in oblique TM-polarization scattering,
as a combination of tangential and normal electric dipoles.
A particular realization is an array of tilted straight
conductor dipole meta-atoms with the dipole axis aligned
with the reflection direction. The geometrical nature of
the synthesized scattering null allows an extremely wide bandwidth of
zero reflection. In the presence of loss in constituent
materials or intended power absorption, it has been
analytically proven and demonstrated using a numerical example
that a complete transmission phase coverage is
possible for reflectionless metasurfaces. The operation
principle and the new design strategy presented in this study will
facilitate development of a new class of transmissive metasurfaces for
wave manipulation with high power efficiencies.

\section*{Acknowledgment}
This work was supported in part by the Academy 
of Finland (project 287894).

%\bibliographystyle{apsrev4-1}
%\nocite{apsrev41Control}
%\bibliography{IEEEabrv,references}

\begin{thebibliography}{45}%
\makeatletter
\providecommand \@ifxundefined [1]{%
 \@ifx{#1\undefined}
}%
\providecommand \@ifnum [1]{%
 \ifnum #1\expandafter \@firstoftwo
 \else \expandafter \@secondoftwo
 \fi
}%
\providecommand \@ifx [1]{%
 \ifx #1\expandafter \@firstoftwo
 \else \expandafter \@secondoftwo
 \fi
}%
\providecommand \natexlab [1]{#1}%
\providecommand \enquote  [1]{``#1''}%
\providecommand \bibnamefont  [1]{#1}%
\providecommand \bibfnamefont [1]{#1}%
\providecommand \citenamefont [1]{#1}%
\providecommand \href@noop [0]{\@secondoftwo}%
\providecommand \href [0]{\begingroup \@sanitize@url \@href}%
\providecommand \@href[1]{\@@startlink{#1}\@@href}%
\providecommand \@@href[1]{\endgroup#1\@@endlink}%
\providecommand \@sanitize@url [0]{\catcode `\\12\catcode `\$12\catcode
  `\&12\catcode `\#12\catcode `\^12\catcode `\_12\catcode `\%12\relax}%
\providecommand \@@startlink[1]{}%
\providecommand \@@endlink[0]{}%
\providecommand \url  [0]{\begingroup\@sanitize@url \@url }%
\providecommand \@url [1]{\endgroup\@href {#1}{\urlprefix }}%
\providecommand \urlprefix  [0]{URL }%
\providecommand \Eprint [0]{\href }%
\providecommand \doibase [0]{http://dx.doi.org/}%
\providecommand \selectlanguage [0]{\@gobble}%
\providecommand \bibinfo  [0]{\@secondoftwo}%
\providecommand \bibfield  [0]{\@secondoftwo}%
\providecommand \translation [1]{[#1]}%
\providecommand \BibitemOpen [0]{}%
\providecommand \bibitemStop [0]{}%
\providecommand \bibitemNoStop [0]{.\EOS\space}%
\providecommand \EOS [0]{\spacefactor3000\relax}%
\providecommand \BibitemShut  [1]{\csname bibitem#1\endcsname}%
\let\auto@bib@innerbib\@empty
%</preamble>
\bibitem [{\citenamefont {Holloway}\ \emph {et~al.}(2012)\citenamefont
  {Holloway}, \citenamefont {Kuester}, \citenamefont {Gordon}, \citenamefont
  {O'Hara}, \citenamefont {Booth},\ and\ \citenamefont
  {Smith}}]{holloway_ieeemap2012}%
  \BibitemOpen
  \bibfield  {author} {\bibinfo {author} {\bibfnamefont {C.~L.}\ \bibnamefont
  {Holloway}}, \bibinfo {author} {\bibfnamefont {E.~F.}\ \bibnamefont
  {Kuester}}, \bibinfo {author} {\bibfnamefont {J.~A.}\ \bibnamefont {Gordon}},
  \bibinfo {author} {\bibfnamefont {J.}~\bibnamefont {O'Hara}}, \bibinfo
  {author} {\bibfnamefont {J.}~\bibnamefont {Booth}}, \ and\ \bibinfo {author}
  {\bibfnamefont {D.~R.}\ \bibnamefont {Smith}},\ }\bibfield  {title} {\enquote
  {\bibinfo {title} {An overview of the theory and applications of
  metasurfaces: the two-dimensional equivalents of metamaterials},}\
  }\href@noop {} {\bibfield  {journal} {\bibinfo  {journal} {{IEEE} Antennas
  Propag. Mag.}\ }\textbf {\bibinfo {volume} {54}},\ \bibinfo {pages} {10}
  (\bibinfo {year} {2012})}\BibitemShut {NoStop}%
\bibitem [{\citenamefont {Kildishev}\ \emph {et~al.}(2013)\citenamefont
  {Kildishev}, \citenamefont {Boltasseva},\ and\ \citenamefont
  {Shalaev}}]{kildishev_science2013}%
  \BibitemOpen
  \bibfield  {author} {\bibinfo {author} {\bibfnamefont {A.~V.}\ \bibnamefont
  {Kildishev}}, \bibinfo {author} {\bibfnamefont {A.}~\bibnamefont
  {Boltasseva}}, \ and\ \bibinfo {author} {\bibfnamefont {V.~M.}\ \bibnamefont
  {Shalaev}},\ }\bibfield  {title} {\enquote {\bibinfo {title} {Planar
  photonics with metasurfaces},}\ }\href@noop {} {\bibfield  {journal}
  {\bibinfo  {journal} {Science}\ }\textbf {\bibinfo {volume} {339}},\ \bibinfo
  {eid} {1232009} (\bibinfo {year} {2013})}\BibitemShut {NoStop}%
\bibitem [{\citenamefont {Yu}\ and\ \citenamefont
  {Capasso}(2014)}]{yu_natmater2014}%
  \BibitemOpen
  \bibfield  {author} {\bibinfo {author} {\bibfnamefont {N.}~\bibnamefont
  {Yu}}\ and\ \bibinfo {author} {\bibfnamefont {F.}~\bibnamefont {Capasso}},\
  }\bibfield  {title} {\enquote {\bibinfo {title} {Flat optics with designer
  metasurfaces},}\ }\href@noop {} {\bibfield  {journal} {\bibinfo  {journal}
  {Nat. Mater.}\ }\textbf {\bibinfo {volume} {13}},\ \bibinfo {pages} {139}
  (\bibinfo {year} {2014})}\BibitemShut {NoStop}%
\bibitem [{\citenamefont {Meinzer}\ \emph {et~al.}(2014)\citenamefont
  {Meinzer}, \citenamefont {Barnes},\ and\ \citenamefont
  {Hooper}}]{meinzer_natphoton2014}%
  \BibitemOpen
  \bibfield  {author} {\bibinfo {author} {\bibfnamefont {N.}~\bibnamefont
  {Meinzer}}, \bibinfo {author} {\bibfnamefont {W.~L.}\ \bibnamefont {Barnes}},
  \ and\ \bibinfo {author} {\bibfnamefont {I.~R.}\ \bibnamefont {Hooper}},\
  }\bibfield  {title} {\enquote {\bibinfo {title} {Plasmonic meta-atoms and
  metasurfaces},}\ }\href@noop {} {\bibfield  {journal} {\bibinfo  {journal}
  {Nat. Photon.}\ }\textbf {\bibinfo {volume} {8}},\ \bibinfo {pages} {889}
  (\bibinfo {year} {2014})}\BibitemShut {NoStop}%
\bibitem [{\citenamefont {Glybovski}\ \emph {et~al.}(2016)\citenamefont
  {Glybovski}, \citenamefont {Tretyakov}, \citenamefont {Belov}, \citenamefont
  {Kivshar},\ and\ \citenamefont {Simovski}}]{glybovski_physrep2016}%
  \BibitemOpen
  \bibfield  {author} {\bibinfo {author} {\bibfnamefont {S.~B.}\ \bibnamefont
  {Glybovski}}, \bibinfo {author} {\bibfnamefont {S.~A.}\ \bibnamefont
  {Tretyakov}}, \bibinfo {author} {\bibfnamefont {P.~A.}\ \bibnamefont
  {Belov}}, \bibinfo {author} {\bibfnamefont {Y.~S.}\ \bibnamefont {Kivshar}},
  \ and\ \bibinfo {author} {\bibfnamefont {C.~R.}\ \bibnamefont {Simovski}},\
  }\bibfield  {title} {\enquote {\bibinfo {title} {Metasurfaces: From
  microwaves to visible},}\ }\href@noop {} {\bibfield  {journal} {\bibinfo
  {journal} {Phys. Rep.}\ }\textbf {\bibinfo {volume} {634}},\ \bibinfo {pages}
  {1} (\bibinfo {year} {2016})}\BibitemShut {NoStop}%
\bibitem [{\citenamefont {Chen}\ \emph {et~al.}(2016)\citenamefont {Chen},
  \citenamefont {Taylor},\ and\ \citenamefont {Yu}}]{chen_repprogphys2016}%
  \BibitemOpen
  \bibfield  {author} {\bibinfo {author} {\bibfnamefont {H.-T.}\ \bibnamefont
  {Chen}}, \bibinfo {author} {\bibfnamefont {A.~J.}\ \bibnamefont {Taylor}}, \
  and\ \bibinfo {author} {\bibfnamefont {N.}~\bibnamefont {Yu}},\ }\bibfield
  {title} {\enquote {\bibinfo {title} {A review of metasurfaces: physics and
  applications},}\ }\href@noop {} {\bibfield  {journal} {\bibinfo  {journal}
  {Rep. Prog. Phys.}\ }\textbf {\bibinfo {volume} {79}},\ \bibinfo {eid}
  {076401} (\bibinfo {year} {2016})}\BibitemShut {NoStop}%
\bibitem [{\citenamefont {Hansen}(2009)}]{hansen2009}%
  \BibitemOpen
  \bibfield  {author} {\bibinfo {author} {\bibfnamefont {R.~C.}\ \bibnamefont
  {Hansen}},\ }\href@noop {} {\emph {\bibinfo {title} {Phased Array
  Antennas}}},\ \bibinfo {edition} {2nd}\ ed.\ (\bibinfo  {publisher} {Wiley},\
  \bibinfo {address} {Hoboken, NJ},\ \bibinfo {year} {2009})\BibitemShut
  {NoStop}%
\bibitem [{\citenamefont {Yu}\ \emph {et~al.}(2011)\citenamefont {Yu},
  \citenamefont {Genevet}, \citenamefont {Kats}, \citenamefont {Aieta},
  \citenamefont {Tetienne}, \citenamefont {Capasso},\ and\ \citenamefont
  {Gaburro}}]{yu_science2011}%
  \BibitemOpen
  \bibfield  {author} {\bibinfo {author} {\bibfnamefont {N.}~\bibnamefont
  {Yu}}, \bibinfo {author} {\bibfnamefont {P.}~\bibnamefont {Genevet}},
  \bibinfo {author} {\bibfnamefont {M.~A.}\ \bibnamefont {Kats}}, \bibinfo
  {author} {\bibfnamefont {F.}~\bibnamefont {Aieta}}, \bibinfo {author}
  {\bibfnamefont {J.-P.}\ \bibnamefont {Tetienne}}, \bibinfo {author}
  {\bibfnamefont {F.}~\bibnamefont {Capasso}}, \ and\ \bibinfo {author}
  {\bibfnamefont {Z.}~\bibnamefont {Gaburro}},\ }\bibfield  {title} {\enquote
  {\bibinfo {title} {Light propagation with phase discontinuities: generalized
  laws of reflection and refraction},}\ }\href@noop {} {\bibfield  {journal}
  {\bibinfo  {journal} {Science}\ }\textbf {\bibinfo {volume} {334}},\ \bibinfo
  {pages} {333} (\bibinfo {year} {2011})}\BibitemShut {NoStop}%
\bibitem [{\citenamefont {Pfeiffer}\ and\ \citenamefont
  {Grbic}(2013{\natexlab{a}})}]{pfeiffer_prl2013}%
  \BibitemOpen
  \bibfield  {author} {\bibinfo {author} {\bibfnamefont {C.}~\bibnamefont
  {Pfeiffer}}\ and\ \bibinfo {author} {\bibfnamefont {A.}~\bibnamefont
  {Grbic}},\ }\bibfield  {title} {\enquote {\bibinfo {title} {Metamaterial
  {Huygens'} surfaces: tailoring wave fronts with reflectionless sheets},}\
  }\href@noop {} {\bibfield  {journal} {\bibinfo  {journal} {Phys. Rev. Lett.}\
  }\textbf {\bibinfo {volume} {110}},\ \bibinfo {eid} {197401} (\bibinfo {year}
  {2013}{\natexlab{a}})}\BibitemShut {NoStop}%
\bibitem [{\citenamefont {Aieta}\ \emph
  {et~al.}(2012{\natexlab{a}})\citenamefont {Aieta}, \citenamefont {Genevet},
  \citenamefont {Yu}, \citenamefont {Kats}, \citenamefont {Gaburro},\ and\
  \citenamefont {Capasso}}]{aieta_nanolett2012b}%
  \BibitemOpen
  \bibfield  {author} {\bibinfo {author} {\bibfnamefont {F.}~\bibnamefont
  {Aieta}}, \bibinfo {author} {\bibfnamefont {P.}~\bibnamefont {Genevet}},
  \bibinfo {author} {\bibfnamefont {N.}~\bibnamefont {Yu}}, \bibinfo {author}
  {\bibfnamefont {M.~A.}\ \bibnamefont {Kats}}, \bibinfo {author}
  {\bibfnamefont {Z.}~\bibnamefont {Gaburro}}, \ and\ \bibinfo {author}
  {\bibfnamefont {F.}~\bibnamefont {Capasso}},\ }\bibfield  {title} {\enquote
  {\bibinfo {title} {Out-of-plane reflection and refraction of light by
  anisotropic optical antenna metasurfaces with phase discontinuities},}\
  }\href@noop {} {\bibfield  {journal} {\bibinfo  {journal} {Nano Lett.}\
  }\textbf {\bibinfo {volume} {12}},\ \bibinfo {pages} {1702} (\bibinfo {year}
  {2012}{\natexlab{a}})}\BibitemShut {NoStop}%
\bibitem [{\citenamefont {Pfeiffer}\ \emph {et~al.}(2014)\citenamefont
  {Pfeiffer}, \citenamefont {Emani}, \citenamefont {Shaltout}, \citenamefont
  {Boltasseva}, \citenamefont {Shalaev},\ and\ \citenamefont
  {Grbic}}]{pfeiffer_nanolett2014}%
  \BibitemOpen
  \bibfield  {author} {\bibinfo {author} {\bibfnamefont {C.}~\bibnamefont
  {Pfeiffer}}, \bibinfo {author} {\bibfnamefont {N.~K.}\ \bibnamefont {Emani}},
  \bibinfo {author} {\bibfnamefont {A.~M.}\ \bibnamefont {Shaltout}}, \bibinfo
  {author} {\bibfnamefont {A.}~\bibnamefont {Boltasseva}}, \bibinfo {author}
  {\bibfnamefont {V.~M.}\ \bibnamefont {Shalaev}}, \ and\ \bibinfo {author}
  {\bibfnamefont {A.}~\bibnamefont {Grbic}},\ }\bibfield  {title} {\enquote
  {\bibinfo {title} {Efficient light bending with isotropic metamaterial
  {Huygens'} surfaces},}\ }\href@noop {} {\bibfield  {journal} {\bibinfo
  {journal} {Nano Lett.}\ }\textbf {\bibinfo {volume} {14}},\ \bibinfo {pages}
  {2491} (\bibinfo {year} {2014})}\BibitemShut {NoStop}%
\bibitem [{\citenamefont {Yu}\ \emph {et~al.}(2015)\citenamefont {Yu},
  \citenamefont {Zhu}, \citenamefont {Paniagua-Dom\'inguez}, \citenamefont
  {Fu}, \citenamefont {Luk'yanchuk},\ and\ \citenamefont
  {Kuznetsov}}]{yu_lasphotonrev2015}%
  \BibitemOpen
  \bibfield  {author} {\bibinfo {author} {\bibfnamefont {Y.~F.}\ \bibnamefont
  {Yu}}, \bibinfo {author} {\bibfnamefont {A.~Y.}\ \bibnamefont {Zhu}},
  \bibinfo {author} {\bibfnamefont {R.}~\bibnamefont {Paniagua-Dom\'inguez}},
  \bibinfo {author} {\bibfnamefont {Y.~H.}\ \bibnamefont {Fu}}, \bibinfo
  {author} {\bibfnamefont {B.}~\bibnamefont {Luk'yanchuk}}, \ and\ \bibinfo
  {author} {\bibfnamefont {A.~I.}\ \bibnamefont {Kuznetsov}},\ }\bibfield
  {title} {\enquote {\bibinfo {title} {High-transmission dielectric metasurface
  with $2\pi$ phase control at visible wavelengths},}\ }\href@noop {}
  {\bibfield  {journal} {\bibinfo  {journal} {Laser Photon. Rev.}\ }\textbf
  {\bibinfo {volume} {9}},\ \bibinfo {pages} {412} (\bibinfo {year}
  {2015})}\BibitemShut {NoStop}%
\bibitem [{\citenamefont {Monticone}\ \emph {et~al.}(2013)\citenamefont
  {Monticone}, \citenamefont {Mohammadi~Estakhri},\ and\ \citenamefont
  {Al\`u}}]{monticone_prl2013}%
  \BibitemOpen
  \bibfield  {author} {\bibinfo {author} {\bibfnamefont {F.}~\bibnamefont
  {Monticone}}, \bibinfo {author} {\bibfnamefont {N.}~\bibnamefont
  {Mohammadi~Estakhri}}, \ and\ \bibinfo {author} {\bibfnamefont
  {A.}~\bibnamefont {Al\`u}},\ }\bibfield  {title} {\enquote {\bibinfo {title}
  {Full control of nanoscale optical transmission with a composite
  metascreen},}\ }\href@noop {} {\bibfield  {journal} {\bibinfo  {journal}
  {Phys. Rev. Lett.}\ }\textbf {\bibinfo {volume} {110}},\ \bibinfo {eid}
  {203903} (\bibinfo {year} {2013})}\BibitemShut {NoStop}%
\bibitem [{\citenamefont {Lin}\ \emph {et~al.}(2014)\citenamefont {Lin},
  \citenamefont {Fan}, \citenamefont {Hasman},\ and\ \citenamefont
  {Brongersma}}]{lin_science2014}%
  \BibitemOpen
  \bibfield  {author} {\bibinfo {author} {\bibfnamefont {D.}~\bibnamefont
  {Lin}}, \bibinfo {author} {\bibfnamefont {P.}~\bibnamefont {Fan}}, \bibinfo
  {author} {\bibfnamefont {E.}~\bibnamefont {Hasman}}, \ and\ \bibinfo {author}
  {\bibfnamefont {M.~L.}\ \bibnamefont {Brongersma}},\ }\bibfield  {title}
  {\enquote {\bibinfo {title} {Dielectric gradient metasurface optical
  elements},}\ }\href@noop {} {\bibfield  {journal} {\bibinfo  {journal}
  {Science}\ }\textbf {\bibinfo {volume} {345}},\ \bibinfo {pages} {298}
  (\bibinfo {year} {2014})}\BibitemShut {NoStop}%
\bibitem [{\citenamefont {Wang}\ \emph {et~al.}(2015)\citenamefont {Wang},
  \citenamefont {Zhang}, \citenamefont {Xu}, \citenamefont {Tian},
  \citenamefont {Gu}, \citenamefont {Yue}, \citenamefont {Zhang}, \citenamefont
  {Han},\ and\ \citenamefont {Zhang}}]{wang_advopticalmat2015}%
  \BibitemOpen
  \bibfield  {author} {\bibinfo {author} {\bibfnamefont {Q.}~\bibnamefont
  {Wang}}, \bibinfo {author} {\bibfnamefont {X.}~\bibnamefont {Zhang}},
  \bibinfo {author} {\bibfnamefont {Y.}~\bibnamefont {Xu}}, \bibinfo {author}
  {\bibfnamefont {Z.}~\bibnamefont {Tian}}, \bibinfo {author} {\bibfnamefont
  {J.}~\bibnamefont {Gu}}, \bibinfo {author} {\bibfnamefont {W.}~\bibnamefont
  {Yue}}, \bibinfo {author} {\bibfnamefont {S.}~\bibnamefont {Zhang}}, \bibinfo
  {author} {\bibfnamefont {J.}~\bibnamefont {Han}}, \ and\ \bibinfo {author}
  {\bibfnamefont {W.}~\bibnamefont {Zhang}},\ }\bibfield  {title} {\enquote
  {\bibinfo {title} {A broadband metasurface-based terahertz flat-lens
  array},}\ }\href@noop {} {\bibfield  {journal} {\bibinfo  {journal} {Adv.
  Opt. Mater.}\ }\textbf {\bibinfo {volume} {3}},\ \bibinfo {pages} {779}
  (\bibinfo {year} {2015})}\BibitemShut {NoStop}%
\bibitem [{\citenamefont {Khorasaninejad}\ \emph {et~al.}(2016)\citenamefont
  {Khorasaninejad}, \citenamefont {Chen}, \citenamefont {Devlin}, \citenamefont
  {Oh}, \citenamefont {Zhu},\ and\ \citenamefont
  {Capasso}}]{khorasaninejad_science2016}%
  \BibitemOpen
  \bibfield  {author} {\bibinfo {author} {\bibfnamefont {M.}~\bibnamefont
  {Khorasaninejad}}, \bibinfo {author} {\bibfnamefont {W.~T.}\ \bibnamefont
  {Chen}}, \bibinfo {author} {\bibfnamefont {R.~C.}\ \bibnamefont {Devlin}},
  \bibinfo {author} {\bibfnamefont {J.}~\bibnamefont {Oh}}, \bibinfo {author}
  {\bibfnamefont {A.~Y.}\ \bibnamefont {Zhu}}, \ and\ \bibinfo {author}
  {\bibfnamefont {F.}~\bibnamefont {Capasso}},\ }\bibfield  {title} {\enquote
  {\bibinfo {title} {Metalenses at visible wavelengths: Diffraction-limited
  focusing and subwavelength resolution imaging},}\ }\href@noop {} {\bibfield
  {journal} {\bibinfo  {journal} {Science}\ }\textbf {\bibinfo {volume}
  {362}},\ \bibinfo {pages} {1190} (\bibinfo {year} {2016})}\BibitemShut
  {NoStop}%
\bibitem [{\citenamefont {Ni}\ \emph {et~al.}(2013)\citenamefont {Ni},
  \citenamefont {Kildishev},\ and\ \citenamefont {Shalaev}}]{ni_natcommun2013}%
  \BibitemOpen
  \bibfield  {author} {\bibinfo {author} {\bibfnamefont {X.}~\bibnamefont
  {Ni}}, \bibinfo {author} {\bibfnamefont {A.~V.}\ \bibnamefont {Kildishev}}, \
  and\ \bibinfo {author} {\bibfnamefont {V.~M.}\ \bibnamefont {Shalaev}},\
  }\bibfield  {title} {\enquote {\bibinfo {title} {Metasurface holograms for
  visible light},}\ }\href@noop {} {\bibfield  {journal} {\bibinfo  {journal}
  {Nat. Commun.}\ }\textbf {\bibinfo {volume} {4}},\ \bibinfo {eid} {2087}
  (\bibinfo {year} {2013})}\BibitemShut {NoStop}%
\bibitem [{\citenamefont {Huang}\ \emph {et~al.}(2013)\citenamefont {Huang},
  \citenamefont {Chen}, \citenamefont {M\"uhlenbernd}, \citenamefont {Zhang},
  \citenamefont {Chen}, \citenamefont {Bai}, \citenamefont {Tan}, \citenamefont
  {Jin}, \citenamefont {Cheah}, \citenamefont {Qiu}, \citenamefont {Li},
  \citenamefont {Zentgraf},\ and\ \citenamefont {Zhang}}]{huang_natcommun2013}%
  \BibitemOpen
  \bibfield  {author} {\bibinfo {author} {\bibfnamefont {L.}~\bibnamefont
  {Huang}}, \bibinfo {author} {\bibfnamefont {X.}~\bibnamefont {Chen}},
  \bibinfo {author} {\bibfnamefont {H.}~\bibnamefont {M\"uhlenbernd}}, \bibinfo
  {author} {\bibfnamefont {H.}~\bibnamefont {Zhang}}, \bibinfo {author}
  {\bibfnamefont {S.}~\bibnamefont {Chen}}, \bibinfo {author} {\bibfnamefont
  {B.}~\bibnamefont {Bai}}, \bibinfo {author} {\bibfnamefont {Q.}~\bibnamefont
  {Tan}}, \bibinfo {author} {\bibfnamefont {G.}~\bibnamefont {Jin}}, \bibinfo
  {author} {\bibfnamefont {K.-W.}\ \bibnamefont {Cheah}}, \bibinfo {author}
  {\bibfnamefont {C.-W.}\ \bibnamefont {Qiu}}, \bibinfo {author} {\bibfnamefont
  {J.}~\bibnamefont {Li}}, \bibinfo {author} {\bibfnamefont {T.}~\bibnamefont
  {Zentgraf}}, \ and\ \bibinfo {author} {\bibfnamefont {S.}~\bibnamefont
  {Zhang}},\ }\bibfield  {title} {\enquote {\bibinfo {title} {Three-dimensional
  optical holography using a plasmonic metasurface},}\ }\href@noop {}
  {\bibfield  {journal} {\bibinfo  {journal} {Nat. Commun.}\ }\textbf {\bibinfo
  {volume} {4}},\ \bibinfo {eid} {2808} (\bibinfo {year} {2013})}\BibitemShut
  {NoStop}%
\bibitem [{\citenamefont {Arbabi}\ \emph {et~al.}(2015)\citenamefont {Arbabi},
  \citenamefont {Horie}, \citenamefont {Bagheri},\ and\ \citenamefont
  {Faraon}}]{arbabi_natnanotech2015}%
  \BibitemOpen
  \bibfield  {author} {\bibinfo {author} {\bibfnamefont {A.}~\bibnamefont
  {Arbabi}}, \bibinfo {author} {\bibfnamefont {Y.}~\bibnamefont {Horie}},
  \bibinfo {author} {\bibfnamefont {M.}~\bibnamefont {Bagheri}}, \ and\
  \bibinfo {author} {\bibfnamefont {A.}~\bibnamefont {Faraon}},\ }\bibfield
  {title} {\enquote {\bibinfo {title} {Dielectric metasurfaces for complete
  control of phase and polarization with subwavelength spatial resolution and
  high transmission},}\ }\href@noop {} {\bibfield  {journal} {\bibinfo
  {journal} {Nat. Nanotechnol.}\ }\textbf {\bibinfo {volume} {10}},\ \bibinfo
  {pages} {937} (\bibinfo {year} {2015})}\BibitemShut {NoStop}%
\bibitem [{\citenamefont {Zhao}\ \emph {et~al.}(2012)\citenamefont {Zhao},
  \citenamefont {Belkin},\ and\ \citenamefont {Al\`u}}]{zhao_natcommun2012}%
  \BibitemOpen
  \bibfield  {author} {\bibinfo {author} {\bibfnamefont {Y.}~\bibnamefont
  {Zhao}}, \bibinfo {author} {\bibfnamefont {M.~A.}\ \bibnamefont {Belkin}}, \
  and\ \bibinfo {author} {\bibfnamefont {A.}~\bibnamefont {Al\`u}},\ }\bibfield
   {title} {\enquote {\bibinfo {title} {Twisted optical metamaterials for
  planarized ultrathin broadband circular polarizers},}\ }\href@noop {}
  {\bibfield  {journal} {\bibinfo  {journal} {Nat. Commun.}\ }\textbf {\bibinfo
  {volume} {3}},\ \bibinfo {eid} {870} (\bibinfo {year} {2012})}\BibitemShut
  {NoStop}%
\bibitem [{\citenamefont {Pfeiffer}\ and\ \citenamefont
  {Grbic}(2013{\natexlab{b}})}]{pfeiffer_ieeejmtt2013}%
  \BibitemOpen
  \bibfield  {author} {\bibinfo {author} {\bibfnamefont {C.}~\bibnamefont
  {Pfeiffer}}\ and\ \bibinfo {author} {\bibfnamefont {A.}~\bibnamefont
  {Grbic}},\ }\bibfield  {title} {\enquote {\bibinfo {title} {Millimeter-wave
  transmitarrays for wavefront and polarization control},}\ }\href@noop {}
  {\bibfield  {journal} {\bibinfo  {journal} {{IEEE} Trans. Microw. Theory
  Techn.}\ }\textbf {\bibinfo {volume} {61}},\ \bibinfo {pages} {4407}
  (\bibinfo {year} {2013}{\natexlab{b}})}\BibitemShut {NoStop}%
\bibitem [{\citenamefont {Grady}\ \emph {et~al.}(2013)\citenamefont {Grady},
  \citenamefont {Heyes}, \citenamefont {Chowdhury}, \citenamefont {Zheng},
  \citenamefont {Reiten}, \citenamefont {Azad}, \citenamefont {Taylor},
  \citenamefont {Dalvit},\ and\ \citenamefont {Chen}}]{grady_science2013}%
  \BibitemOpen
  \bibfield  {author} {\bibinfo {author} {\bibfnamefont {N.~K.}\ \bibnamefont
  {Grady}}, \bibinfo {author} {\bibfnamefont {J.~E.}\ \bibnamefont {Heyes}},
  \bibinfo {author} {\bibfnamefont {D.~R.}\ \bibnamefont {Chowdhury}}, \bibinfo
  {author} {\bibfnamefont {Y.}~\bibnamefont {Zheng}}, \bibinfo {author}
  {\bibfnamefont {M.~T.}\ \bibnamefont {Reiten}}, \bibinfo {author}
  {\bibfnamefont {A.~K.}\ \bibnamefont {Azad}}, \bibinfo {author}
  {\bibfnamefont {A.~J.}\ \bibnamefont {Taylor}}, \bibinfo {author}
  {\bibfnamefont {D.~A.~R.}\ \bibnamefont {Dalvit}}, \ and\ \bibinfo {author}
  {\bibfnamefont {H.-T.}\ \bibnamefont {Chen}},\ }\bibfield  {title} {\enquote
  {\bibinfo {title} {Terahertz metamaterials for linear polarization conversion
  and anomalous refraction},}\ }\href@noop {} {\bibfield  {journal} {\bibinfo
  {journal} {Science}\ }\textbf {\bibinfo {volume} {340}},\ \bibinfo {pages}
  {1304} (\bibinfo {year} {2013})}\BibitemShut {NoStop}%
\bibitem [{\citenamefont {Aieta}\ \emph
  {et~al.}(2012{\natexlab{b}})\citenamefont {Aieta}, \citenamefont {Genevet},
  \citenamefont {Kats}, \citenamefont {Yu}, \citenamefont {Blanchard},
  \citenamefont {Gaburro},\ and\ \citenamefont {Capasso}}]{aieta_nanolett2012}%
  \BibitemOpen
  \bibfield  {author} {\bibinfo {author} {\bibfnamefont {F.}~\bibnamefont
  {Aieta}}, \bibinfo {author} {\bibfnamefont {P.}~\bibnamefont {Genevet}},
  \bibinfo {author} {\bibfnamefont {M.~A.}\ \bibnamefont {Kats}}, \bibinfo
  {author} {\bibfnamefont {N.}~\bibnamefont {Yu}}, \bibinfo {author}
  {\bibfnamefont {R.}~\bibnamefont {Blanchard}}, \bibinfo {author}
  {\bibfnamefont {Z.}~\bibnamefont {Gaburro}}, \ and\ \bibinfo {author}
  {\bibfnamefont {F.}~\bibnamefont {Capasso}},\ }\bibfield  {title} {\enquote
  {\bibinfo {title} {Aberration-free ultrathin flat lenses and axicons at
  telecom wavelengths based on plasmonic metasurfaces},}\ }\href@noop {}
  {\bibfield  {journal} {\bibinfo  {journal} {Nano Lett.}\ }\textbf {\bibinfo
  {volume} {12}},\ \bibinfo {pages} {4932} (\bibinfo {year}
  {2012}{\natexlab{b}})}\BibitemShut {NoStop}%
\bibitem [{\citenamefont {Zheng}\ \emph {et~al.}(2015)\citenamefont {Zheng},
  \citenamefont {M\"uhlenbernd}, \citenamefont {Kenney}, \citenamefont {Li},
  \citenamefont {Zentgraf},\ and\ \citenamefont
  {Zhang}}]{zheng_natnanotech2015}%
  \BibitemOpen
  \bibfield  {author} {\bibinfo {author} {\bibfnamefont {G.}~\bibnamefont
  {Zheng}}, \bibinfo {author} {\bibfnamefont {H.}~\bibnamefont
  {M\"uhlenbernd}}, \bibinfo {author} {\bibfnamefont {M.}~\bibnamefont
  {Kenney}}, \bibinfo {author} {\bibfnamefont {G.}~\bibnamefont {Li}}, \bibinfo
  {author} {\bibfnamefont {T.}~\bibnamefont {Zentgraf}}, \ and\ \bibinfo
  {author} {\bibfnamefont {S.}~\bibnamefont {Zhang}},\ }\bibfield  {title}
  {\enquote {\bibinfo {title} {Metasurface holograms reaching 80\%
  efficiency},}\ }\href@noop {} {\bibfield  {journal} {\bibinfo  {journal}
  {Nat. Nanotech.}\ }\textbf {\bibinfo {volume} {10}},\ \bibinfo {pages} {308}
  (\bibinfo {year} {2015})}\BibitemShut {NoStop}%
\bibitem [{\citenamefont {Shalaev}\ \emph {et~al.}(2015)\citenamefont
  {Shalaev}, \citenamefont {Sun}, \citenamefont {Tsukernik}, \citenamefont
  {Pandey}, \citenamefont {Nikolskiy},\ and\ \citenamefont
  {Litchinitser}}]{shalaev_nanolett2015}%
  \BibitemOpen
  \bibfield  {author} {\bibinfo {author} {\bibfnamefont {M.~I.}\ \bibnamefont
  {Shalaev}}, \bibinfo {author} {\bibfnamefont {J.}~\bibnamefont {Sun}},
  \bibinfo {author} {\bibfnamefont {A.}~\bibnamefont {Tsukernik}}, \bibinfo
  {author} {\bibfnamefont {A.}~\bibnamefont {Pandey}}, \bibinfo {author}
  {\bibfnamefont {K.}~\bibnamefont {Nikolskiy}}, \ and\ \bibinfo {author}
  {\bibfnamefont {N.~M.}\ \bibnamefont {Litchinitser}},\ }\bibfield  {title}
  {\enquote {\bibinfo {title} {High-efficiency all-dielectric metasurfaces for
  ultracompact beam manipulation in transmission mode},}\ }\href@noop {}
  {\bibfield  {journal} {\bibinfo  {journal} {Nano Lett.}\ }\textbf {\bibinfo
  {volume} {15}},\ \bibinfo {pages} {6261} (\bibinfo {year}
  {2015})}\BibitemShut {NoStop}%
\bibitem [{\citenamefont {Decker}\ \emph {et~al.}(2015)\citenamefont {Decker},
  \citenamefont {Staude}, \citenamefont {Falkner}, \citenamefont {Dominguez},
  \citenamefont {Neshev}, \citenamefont {Brener}, \citenamefont {Pertsch},\
  and\ \citenamefont {Kivshar}}]{decker_advoptmat2015}%
  \BibitemOpen
  \bibfield  {author} {\bibinfo {author} {\bibfnamefont {M.}~\bibnamefont
  {Decker}}, \bibinfo {author} {\bibfnamefont {I.}~\bibnamefont {Staude}},
  \bibinfo {author} {\bibfnamefont {M.}~\bibnamefont {Falkner}}, \bibinfo
  {author} {\bibfnamefont {J.}~\bibnamefont {Dominguez}}, \bibinfo {author}
  {\bibfnamefont {D.~N.}\ \bibnamefont {Neshev}}, \bibinfo {author}
  {\bibfnamefont {I.}~\bibnamefont {Brener}}, \bibinfo {author} {\bibfnamefont
  {T.}~\bibnamefont {Pertsch}}, \ and\ \bibinfo {author} {\bibfnamefont
  {Y.~S.}\ \bibnamefont {Kivshar}},\ }\bibfield  {title} {\enquote {\bibinfo
  {title} {High-efficiency dielectric {Huygens'} surfaces},}\ }\href@noop {}
  {\bibfield  {journal} {\bibinfo  {journal} {Adv. Optical Mater.}\ }\textbf
  {\bibinfo {volume} {3}},\ \bibinfo {pages} {813} (\bibinfo {year}
  {2015})}\BibitemShut {NoStop}%
\bibitem [{\citenamefont {Chong}\ \emph {et~al.}(2016)\citenamefont {Chong},
  \citenamefont {Wang}, \citenamefont {Staude}, \citenamefont {James},
  \citenamefont {Dominguez}, \citenamefont {Liu}, \citenamefont {Subramania},
  \citenamefont {Decker}, \citenamefont {Neshev}, \citenamefont {Brener},\ and\
  \citenamefont {Kivshar}}]{chong_acsphoton2016}%
  \BibitemOpen
  \bibfield  {author} {\bibinfo {author} {\bibfnamefont {K.~E.}\ \bibnamefont
  {Chong}}, \bibinfo {author} {\bibfnamefont {L.}~\bibnamefont {Wang}},
  \bibinfo {author} {\bibfnamefont {I.}~\bibnamefont {Staude}}, \bibinfo
  {author} {\bibfnamefont {A.~R.}\ \bibnamefont {James}}, \bibinfo {author}
  {\bibfnamefont {J.}~\bibnamefont {Dominguez}}, \bibinfo {author}
  {\bibfnamefont {S.}~\bibnamefont {Liu}}, \bibinfo {author} {\bibfnamefont
  {G.~S.}\ \bibnamefont {Subramania}}, \bibinfo {author} {\bibfnamefont
  {M.}~\bibnamefont {Decker}}, \bibinfo {author} {\bibfnamefont {D.~N.}\
  \bibnamefont {Neshev}}, \bibinfo {author} {\bibfnamefont {I.}~\bibnamefont
  {Brener}}, \ and\ \bibinfo {author} {\bibfnamefont {Y.~S.}\ \bibnamefont
  {Kivshar}},\ }\bibfield  {title} {\enquote {\bibinfo {title} {Efficient
  polarization-insensitive complex wavefront control using {Huygens'}
  metasurfaces based on dielectric resonant meta-atoms},}\ }\href@noop {}
  {\bibfield  {journal} {\bibinfo  {journal} {ACS Photon.}\ }\textbf {\bibinfo
  {volume} {3}},\ \bibinfo {pages} {514} (\bibinfo {year} {2016})}\BibitemShut
  {NoStop}%
\bibitem [{\citenamefont {Wong}\ \emph {et~al.}(2015)\citenamefont {Wong},
  \citenamefont {Selvanayagam},\ and\ \citenamefont
  {Eleftheriades}}]{wong_ieeejmtt2015}%
  \BibitemOpen
  \bibfield  {author} {\bibinfo {author} {\bibfnamefont {J.~P.~S.}\
  \bibnamefont {Wong}}, \bibinfo {author} {\bibfnamefont {M.}~\bibnamefont
  {Selvanayagam}}, \ and\ \bibinfo {author} {\bibfnamefont {G.~V.}\
  \bibnamefont {Eleftheriades}},\ }\bibfield  {title} {\enquote {\bibinfo
  {title} {Polarization considerations for scalar {Huygens} metasurfaces and
  characterization for {2-D} refraction},}\ }\href@noop {} {\bibfield
  {journal} {\bibinfo  {journal} {{IEEE} Trans. Microw. Theory Techn.}\
  }\textbf {\bibinfo {volume} {63}},\ \bibinfo {pages} {913} (\bibinfo {year}
  {2015})}\BibitemShut {NoStop}%
\bibitem [{\citenamefont {Epstein}\ and\ \citenamefont
  {Eleftheriades}(2016)}]{epstein_josab2016}%
  \BibitemOpen
  \bibfield  {author} {\bibinfo {author} {\bibfnamefont {A.}~\bibnamefont
  {Epstein}}\ and\ \bibinfo {author} {\bibfnamefont {G.~V.}\ \bibnamefont
  {Eleftheriades}},\ }\bibfield  {title} {\enquote {\bibinfo {title}
  {{Huygens'} metasurfaces via the equivalence principle: design and
  applications},}\ }\href@noop {} {\bibfield  {journal} {\bibinfo  {journal}
  {J. Opt. Soc. Am. B}\ }\textbf {\bibinfo {volume} {33}},\ \bibinfo {pages}
  {A31} (\bibinfo {year} {2016})}\BibitemShut {NoStop}%
\bibitem [{\citenamefont {Cai}\ \emph {et~al.}(2017{\natexlab{a}})\citenamefont
  {Cai}, \citenamefont {Tang}, \citenamefont {Wang}, \citenamefont {Xu},
  \citenamefont {Sun}, \citenamefont {He},\ and\ \citenamefont
  {Zhou}}]{cai_advoptmat2017}%
  \BibitemOpen
  \bibfield  {author} {\bibinfo {author} {\bibfnamefont {T.}~\bibnamefont
  {Cai}}, \bibinfo {author} {\bibfnamefont {S.}~\bibnamefont {Tang}}, \bibinfo
  {author} {\bibfnamefont {G.}~\bibnamefont {Wang}}, \bibinfo {author}
  {\bibfnamefont {H.}~\bibnamefont {Xu}}, \bibinfo {author} {\bibfnamefont
  {S.}~\bibnamefont {Sun}}, \bibinfo {author} {\bibfnamefont {Q.}~\bibnamefont
  {He}}, \ and\ \bibinfo {author} {\bibfnamefont {L.}~\bibnamefont {Zhou}},\
  }\bibfield  {title} {\enquote {\bibinfo {title} {High-performance
  bifunctional metasurfaces in transmission and reflection geometries},}\
  }\href@noop {} {\bibfield  {journal} {\bibinfo  {journal} {Adv. Opt. Mater.}\
  }\textbf {\bibinfo {volume} {5}},\ \bibinfo {eid} {1600506} (\bibinfo {year}
  {2017}{\natexlab{a}})}\BibitemShut {NoStop}%
\bibitem [{\citenamefont {Luo}\ \emph {et~al.}(2017)\citenamefont {Luo},
  \citenamefont {Sun}, \citenamefont {Xu}, \citenamefont {He},\ and\
  \citenamefont {Zhou}}]{luo_prappl2017}%
  \BibitemOpen
  \bibfield  {author} {\bibinfo {author} {\bibfnamefont {W.}~\bibnamefont
  {Luo}}, \bibinfo {author} {\bibfnamefont {S.}~\bibnamefont {Sun}}, \bibinfo
  {author} {\bibfnamefont {H.-X.}\ \bibnamefont {Xu}}, \bibinfo {author}
  {\bibfnamefont {Q.}~\bibnamefont {He}}, \ and\ \bibinfo {author}
  {\bibfnamefont {L.}~\bibnamefont {Zhou}},\ }\bibfield  {title} {\enquote
  {\bibinfo {title} {Transmissive ultrathin {Pancharatnam-Berry} metasurfaces
  with nearly 100\% efficiency},}\ }\href@noop {} {\bibfield  {journal}
  {\bibinfo  {journal} {Phys. Rev. Appl.}\ }\textbf {\bibinfo {volume} {7}},\
  \bibinfo {eid} {044033} (\bibinfo {year} {2017})}\BibitemShut {NoStop}%
\bibitem [{\citenamefont {Cai}\ \emph {et~al.}(2017{\natexlab{b}})\citenamefont
  {Cai}, \citenamefont {Wang}, \citenamefont {Tang}, \citenamefont {Xu},
  \citenamefont {Duan}, \citenamefont {Guo}, \citenamefont {Guan},
  \citenamefont {He},\ and\ \citenamefont {Zhou}}]{cai_prappl2017}%
  \BibitemOpen
  \bibfield  {author} {\bibinfo {author} {\bibfnamefont {T.}~\bibnamefont
  {Cai}}, \bibinfo {author} {\bibfnamefont {G.}~\bibnamefont {Wang}}, \bibinfo
  {author} {\bibfnamefont {S.}~\bibnamefont {Tang}}, \bibinfo {author}
  {\bibfnamefont {H.}~\bibnamefont {Xu}}, \bibinfo {author} {\bibfnamefont
  {J.}~\bibnamefont {Duan}}, \bibinfo {author} {\bibfnamefont {H.}~\bibnamefont
  {Guo}}, \bibinfo {author} {\bibfnamefont {F.}~\bibnamefont {Guan}}, \bibinfo
  {author} {\bibfnamefont {S.~S.~Q.}\ \bibnamefont {He}}, \ and\ \bibinfo
  {author} {\bibfnamefont {L.}~\bibnamefont {Zhou}},\ }\bibfield  {title}
  {\enquote {\bibinfo {title} {High-efficiency and full-space manipulation of
  electromagnetic wave fronts with metasurfaces},}\ }\href@noop {} {\bibfield
  {journal} {\bibinfo  {journal} {Phys. Rev. Appl.}\ }\textbf {\bibinfo
  {volume} {8}},\ \bibinfo {eid} {034033} (\bibinfo {year}
  {2017}{\natexlab{b}})}\BibitemShut {NoStop}%
\bibitem [{\citenamefont {Niemi}\ \emph {et~al.}(2013)\citenamefont {Niemi},
  \citenamefont {Karilainen},\ and\ \citenamefont
  {Tretyakov}}]{niemi_ieeejap2013}%
  \BibitemOpen
  \bibfield  {author} {\bibinfo {author} {\bibfnamefont {T.}~\bibnamefont
  {Niemi}}, \bibinfo {author} {\bibfnamefont {A.~O.}\ \bibnamefont
  {Karilainen}}, \ and\ \bibinfo {author} {\bibfnamefont {S.~A.}\ \bibnamefont
  {Tretyakov}},\ }\bibfield  {title} {\enquote {\bibinfo {title} {Synthesis of
  polarization transformers},}\ }\href@noop {} {\bibfield  {journal} {\bibinfo
  {journal} {{IEEE} Trans. Antennas Propag.}\ }\textbf {\bibinfo {volume}
  {61}},\ \bibinfo {pages} {3102} (\bibinfo {year} {2013})}\BibitemShut
  {NoStop}%
\bibitem [{\citenamefont {Albooyeh}\ \emph {et~al.}(2017)\citenamefont
  {Albooyeh}, \citenamefont {Kwon}, \citenamefont {Capolino},\ and\
  \citenamefont {Tretyakov}}]{albooyeh_prb2017}%
  \BibitemOpen
  \bibfield  {author} {\bibinfo {author} {\bibfnamefont {M.}~\bibnamefont
  {Albooyeh}}, \bibinfo {author} {\bibfnamefont {D.-H.}\ \bibnamefont {Kwon}},
  \bibinfo {author} {\bibfnamefont {F.}~\bibnamefont {Capolino}}, \ and\
  \bibinfo {author} {\bibfnamefont {S.~A.}\ \bibnamefont {Tretyakov}},\
  }\bibfield  {title} {\enquote {\bibinfo {title} {Equivalent realizations of
  reciprocal metasurfaces: role of tangential and normal polarization},}\
  }\href@noop {} {\bibfield  {journal} {\bibinfo  {journal} {Phys. Rev. B}\
  }\textbf {\bibinfo {volume} {95}},\ \bibinfo {eid} {115435} (\bibinfo {year}
  {2017})}\BibitemShut {NoStop}%
\bibitem [{\citenamefont {Huang}\ \emph {et~al.}(2010)\citenamefont {Huang},
  \citenamefont {Peng},\ and\ \citenamefont {Fan}}]{huang_prl2010}%
  \BibitemOpen
  \bibfield  {author} {\bibinfo {author} {\bibfnamefont {X.-R.}\ \bibnamefont
  {Huang}}, \bibinfo {author} {\bibfnamefont {R.-W.}\ \bibnamefont {Peng}}, \
  and\ \bibinfo {author} {\bibfnamefont {R.-H.}\ \bibnamefont {Fan}},\
  }\bibfield  {title} {\enquote {\bibinfo {title} {Making metals transparent
  for white light by spoof surface plasmons},}\ }\href@noop {} {\bibfield
  {journal} {\bibinfo  {journal} {Phys. Rev. Lett.}\ }\textbf {\bibinfo
  {volume} {105}},\ \bibinfo {eid} {243901} (\bibinfo {year}
  {2010})}\BibitemShut {NoStop}%
\bibitem [{\citenamefont {Al\`u}\ \emph {et~al.}(2011)\citenamefont {Al\`u},
  \citenamefont {D'Aguanno}, \citenamefont {Mattiucci},\ and\ \citenamefont
  {Bloemer}}]{alu_prl2011}%
  \BibitemOpen
  \bibfield  {author} {\bibinfo {author} {\bibfnamefont {A.}~\bibnamefont
  {Al\`u}}, \bibinfo {author} {\bibfnamefont {G.}~\bibnamefont {D'Aguanno}},
  \bibinfo {author} {\bibfnamefont {N.}~\bibnamefont {Mattiucci}}, \ and\
  \bibinfo {author} {\bibfnamefont {M.~J.}\ \bibnamefont {Bloemer}},\
  }\bibfield  {title} {\enquote {\bibinfo {title} {Plasmonic {Brewster} angle:
  broadband extraordinary transmission through optical gratings},}\ }\href@noop
  {} {\bibfield  {journal} {\bibinfo  {journal} {Phys. Rev. Lett.}\ }\textbf
  {\bibinfo {volume} {106}},\ \bibinfo {eid} {123902} (\bibinfo {year}
  {2011})}\BibitemShut {NoStop}%
\bibitem [{\citenamefont {Fan}\ \emph {et~al.}(2012)\citenamefont {Fan},
  \citenamefont {Peng}, \citenamefont {Huang}, \citenamefont {Li},
  \citenamefont {Liu}, \citenamefont {Hu}, \citenamefont {Wang},\ and\
  \citenamefont {Zhang}}]{fan_advmat2012}%
  \BibitemOpen
  \bibfield  {author} {\bibinfo {author} {\bibfnamefont {R.-H.}\ \bibnamefont
  {Fan}}, \bibinfo {author} {\bibfnamefont {R.-W.}\ \bibnamefont {Peng}},
  \bibinfo {author} {\bibfnamefont {X.-R.}\ \bibnamefont {Huang}}, \bibinfo
  {author} {\bibfnamefont {J.}~\bibnamefont {Li}}, \bibinfo {author}
  {\bibfnamefont {Y.}~\bibnamefont {Liu}}, \bibinfo {author} {\bibfnamefont
  {Q.}~\bibnamefont {Hu}}, \bibinfo {author} {\bibfnamefont {M.}~\bibnamefont
  {Wang}}, \ and\ \bibinfo {author} {\bibfnamefont {X.}~\bibnamefont {Zhang}},\
  }\bibfield  {title} {\enquote {\bibinfo {title} {Transparent metals for
  ultrabroadband electromagnetic waves},}\ }\href@noop {} {\bibfield  {journal}
  {\bibinfo  {journal} {Adv. Mater.}\ }\textbf {\bibinfo {volume} {24}},\
  \bibinfo {pages} {1980} (\bibinfo {year} {2012})}\BibitemShut {NoStop}%
\bibitem [{\citenamefont {Bohren}\ and\ \citenamefont
  {Huffman}(2004)}]{bohren2004}%
  \BibitemOpen
  \bibfield  {author} {\bibinfo {author} {\bibfnamefont {C.~F.}\ \bibnamefont
  {Bohren}}\ and\ \bibinfo {author} {\bibfnamefont {D.~R.}\ \bibnamefont
  {Huffman}},\ }\href@noop {} {\emph {\bibinfo {title} {Absorption and
  Scattering of Light by Small Particles}}}\ (\bibinfo  {publisher}
  {Wiley-VCH},\ \bibinfo {address} {Weinheim, Germany},\ \bibinfo {year}
  {2004})\BibitemShut {NoStop}%
\bibitem [{\citenamefont {Asadchy}\ \emph {et~al.}(2015)\citenamefont
  {Asadchy}, \citenamefont {Faniayeu}, \citenamefont {Ra'di}, \citenamefont
  {Khakhomov}, \citenamefont {Semchenko},\ and\ \citenamefont
  {Tretyakov}}]{asadchy_prx2015}%
  \BibitemOpen
  \bibfield  {author} {\bibinfo {author} {\bibfnamefont {V.~S.}\ \bibnamefont
  {Asadchy}}, \bibinfo {author} {\bibfnamefont {I.~A.}\ \bibnamefont
  {Faniayeu}}, \bibinfo {author} {\bibfnamefont {Y.}~\bibnamefont {Ra'di}},
  \bibinfo {author} {\bibfnamefont {S.~A.}\ \bibnamefont {Khakhomov}}, \bibinfo
  {author} {\bibfnamefont {I.~V.}\ \bibnamefont {Semchenko}}, \ and\ \bibinfo
  {author} {\bibfnamefont {S.~A.}\ \bibnamefont {Tretyakov}},\ }\bibfield
  {title} {\enquote {\bibinfo {title} {Broadband reflectionless metasheets:
  frequency-selective transmission and perfect absorption},}\ }\href@noop {}
  {\bibfield  {journal} {\bibinfo  {journal} {Phys. Rev. X}\ }\textbf {\bibinfo
  {volume} {5}},\ \bibinfo {eid} {031005} (\bibinfo {year} {2015})}\BibitemShut
  {NoStop}%
\bibitem [{\citenamefont {Paniagua-Dom\'inguez}\ \emph
  {et~al.}(2016)\citenamefont {Paniagua-Dom\'inguez}, \citenamefont {Yu},
  \citenamefont {Miroshnichenko}, \citenamefont {Krivitsky}, \citenamefont
  {Fu}, \citenamefont {Valuckas}, \citenamefont {Gonzaga}, \citenamefont {Toh},
  \citenamefont {Kay}, \citenamefont {Luk'yanchuk},\ and\ \citenamefont
  {Kuznetsov}}]{paniagua-dominguez_natcommun2016}%
  \BibitemOpen
  \bibfield  {author} {\bibinfo {author} {\bibfnamefont {R.}~\bibnamefont
  {Paniagua-Dom\'inguez}}, \bibinfo {author} {\bibfnamefont {Y.~F.}\
  \bibnamefont {Yu}}, \bibinfo {author} {\bibfnamefont {A.~E.}\ \bibnamefont
  {Miroshnichenko}}, \bibinfo {author} {\bibfnamefont {L.~A.}\ \bibnamefont
  {Krivitsky}}, \bibinfo {author} {\bibfnamefont {Y.~H.}\ \bibnamefont {Fu}},
  \bibinfo {author} {\bibfnamefont {V.}~\bibnamefont {Valuckas}}, \bibinfo
  {author} {\bibfnamefont {L.}~\bibnamefont {Gonzaga}}, \bibinfo {author}
  {\bibfnamefont {Y.~T.}\ \bibnamefont {Toh}}, \bibinfo {author} {\bibfnamefont
  {A.~Y.~S.}\ \bibnamefont {Kay}}, \bibinfo {author} {\bibfnamefont
  {B.}~\bibnamefont {Luk'yanchuk}}, \ and\ \bibinfo {author} {\bibfnamefont
  {A.~I.}\ \bibnamefont {Kuznetsov}},\ }\bibfield  {title} {\enquote {\bibinfo
  {title} {Generalized {Brewster} effect in dielectric metasurfaces},}\
  }\href@noop {} {\bibfield  {journal} {\bibinfo  {journal} {Nat. Commun.}\
  }\textbf {\bibinfo {volume} {7}},\ \bibinfo {eid} {10362} (\bibinfo {year}
  {2016})}\BibitemShut {NoStop}%
\bibitem [{\citenamefont {Green}(1966)}]{green_ieeejap1966}%
  \BibitemOpen
  \bibfield  {author} {\bibinfo {author} {\bibfnamefont {R.~B.}\ \bibnamefont
  {Green}},\ }\bibfield  {title} {\enquote {\bibinfo {title} {Scattering from
  conjugate-matched antennas},}\ }\href@noop {} {\bibfield  {journal} {\bibinfo
   {journal} {{IEEE} Trans. Antennas Propag.}\ }\textbf {\bibinfo {volume}
  {AP-14}},\ \bibinfo {pages} {17} (\bibinfo {year} {1966})}\BibitemShut
  {NoStop}%
\bibitem [{\citenamefont {Jin}\ and\ \citenamefont
  {Ziolkowski}(2010)}]{jin_ieeejawpl2010}%
  \BibitemOpen
  \bibfield  {author} {\bibinfo {author} {\bibfnamefont {P.}~\bibnamefont
  {Jin}}\ and\ \bibinfo {author} {\bibfnamefont {R.~W.}\ \bibnamefont
  {Ziolkowski}},\ }\bibfield  {title} {\enquote {\bibinfo {title}
  {Metamaterial-inspired, electrically small {Huygens} sources},}\ }\href@noop
  {} {\bibfield  {journal} {\bibinfo  {journal} {{IEEE} Antennas Wireless
  Propag. Lett.}\ }\textbf {\bibinfo {volume} {9}},\ \bibinfo {pages} {501}
  (\bibinfo {year} {2010})}\BibitemShut {NoStop}%
\bibitem [{\citenamefont {Ra'di}\ \emph {et~al.}(2015)\citenamefont {Ra'di},
  \citenamefont {Simovski},\ and\ \citenamefont {Tretyakov}}]{radi_prappl2015}%
  \BibitemOpen
  \bibfield  {author} {\bibinfo {author} {\bibfnamefont {Y.}~\bibnamefont
  {Ra'di}}, \bibinfo {author} {\bibfnamefont {C.~R.}\ \bibnamefont {Simovski}},
  \ and\ \bibinfo {author} {\bibfnamefont {S.~A.}\ \bibnamefont {Tretyakov}},\
  }\bibfield  {title} {\enquote {\bibinfo {title} {Thin perfect absorbers for
  electromagnetic waves: theory, design, and realizations},}\ }\href@noop {}
  {\bibfield  {journal} {\bibinfo  {journal} {Phys. Rev. Appl.}\ }\textbf
  {\bibinfo {volume} {3}},\ \bibinfo {eid} {037001} (\bibinfo {year}
  {2015})}\BibitemShut {NoStop}%
\bibitem [{\citenamefont {Wang}\ \emph {et~al.}(2017)\citenamefont {Wang},
  \citenamefont {D\'iaz-Rubio},\ and\ \citenamefont
  {Tretyakov}}]{wang_ieeejmtt2017}%
  \BibitemOpen
  \bibfield  {author} {\bibinfo {author} {\bibfnamefont {X.-C.}\ \bibnamefont
  {Wang}}, \bibinfo {author} {\bibfnamefont {A.}~\bibnamefont {D\'iaz-Rubio}},
  \ and\ \bibinfo {author} {\bibfnamefont {S.~A.}\ \bibnamefont {Tretyakov}},\
  }\bibfield  {title} {\enquote {\bibinfo {title} {An accurate method for
  measuring the sheet impedance of thin conductive films at microwave and
  millimeter-wave frequencies},}\ }\href@noop {} {\bibfield  {journal}
  {\bibinfo  {journal} {{IEEE} Trans. Microw. Theory Techn.}\ }\textbf
  {\bibinfo {volume} {65}},\ \bibinfo {pages} {5009} (\bibinfo {year}
  {2017})}\BibitemShut {NoStop}%
\bibitem [{\citenamefont {Monti}\ \emph {et~al.}(2016)\citenamefont {Monti},
  \citenamefont {Toscano},\ and\ \citenamefont {Bilotti}}]{monti_oplett2016}%
  \BibitemOpen
  \bibfield  {author} {\bibinfo {author} {\bibfnamefont {A.}~\bibnamefont
  {Monti}}, \bibinfo {author} {\bibfnamefont {A.}~\bibnamefont {Toscano}}, \
  and\ \bibinfo {author} {\bibfnamefont {F.}~\bibnamefont {Bilotti}},\
  }\bibfield  {title} {\enquote {\bibinfo {title} {Exploiting the surface
  dispersion of nanoparticles to design optical-resistive sheets and
  {Salisbury} absorbers},}\ }\href@noop {} {\bibfield  {journal} {\bibinfo
  {journal} {Opt. Lett.}\ }\textbf {\bibinfo {volume} {41}},\ \bibinfo {pages}
  {3383} (\bibinfo {year} {2016})}\BibitemShut {NoStop}%
\end{thebibliography}%


\end{document}